\documentclass[final,superscriptaddress,english,twocolumn,amssymb, nobibnotes, aps, pre, longbibliography]{revtex4-1}
\usepackage{graphicx}

\usepackage{natbib}
\usepackage{amsmath}
\usepackage{amssymb}
\usepackage{appendix}
\usepackage{comment}
\usepackage[dvipsnames]{xcolor}
\usepackage[section]{placeins}
\definecolor{darkblue}{rgb}{0,0,.65}
\definecolor{darkgreen}{rgb}{0.28,0.41,0.19}

\usepackage[%
    pdfauthor={David J. Luitz},%
  pdfstartview=FitH,%
  breaklinks=true,%
  bookmarks=true,%
  colorlinks=true,%
  anchorcolor=black,%
  citecolor=blue,
  filecolor=black,%
  menucolor=black,%
  urlcolor=darkblue,%
  linkcolor=blue,%
 ]{hyperref}
\usepackage[all]{hypcap} 
\usepackage{braket}

\newcommand{\tr}{\mathrm{Tr}}

\newcommand{\E}{\mathrm{e}}

\newcommand{\D}{\mathrm{d}}

\begin{document}

\title{Statistics of correlation functions in the random Heisenberg chain}%

\author{Luis Colmenarez}
\affiliation{Max Planck Institute for the Physics of Complex Systems, Noethnitzer Str. 38, Dresden, Germany}
\author{Paul A. McClarty}
\affiliation{Max Planck Institute for the Physics of Complex Systems, Noethnitzer Str. 38, Dresden, Germany}
\author{Masudul Haque}
\affiliation{Max Planck Institute for the Physics of Complex Systems, Noethnitzer Str. 38, Dresden, Germany}
\affiliation{Department of Theoretical Physics, Maynooth University, Co. Kildare, Ireland}
\author{David J. Luitz}
\affiliation{Max Planck Institute for the Physics of Complex Systems, Noethnitzer Str. 38, Dresden, Germany}
\email{dluitz@pks.mpg.de}
\date{\today}%

\begin{abstract}
Ergodic quantum many-body systems satisfy the eigenstate thermalization hypothesis (ETH). However, strong disorder can destroy ergodicity through many-body localization (MBL) -- at least in one dimensional systems -- leading to a clear signal of the MBL transition in the probability distributions of energy eigenstate expectation values of local operators. For a paradigmatic model of MBL, namely the random-field Heisenberg spin chain, we consider the full probability distribution of \emph{eigenstate correlation functions} across the entire phase diagram. We find gaussian distributions at weak disorder, as predicted by pure ETH. At intermediate disorder -- in the thermal phase -- we find further evidence for \emph{anomalous thermalization} in the form of heavy tails of the distributions. In the MBL phase, we observe peculiar features of the correlator distributions: a strong asymmetry in $S_i^z S_{i+r}^z$ correlators skewed towards negative values; and a multimodal distribution for spin-flip correlators. A \emph{quantitative quasi-degenerate perturbation theory} calculation of these correlators yields a surprising \emph{agreement of the full distribution} with the exact results, revealing, in particular, the origin of the multiple peaks in the spin-flip correlator distribution as arising from the resonant and off-resonant admixture of spin configurations.  The distribution of the $S_i^zS_{i+r}^z$ correlator exhibits striking differences between the MBL and Anderson insulator cases. 
\end{abstract}
\maketitle

\section{Introduction}

Over the last decade, the Eigenstate Thermalization Hypothesis (ETH)
\cite{peres_ergodicity_1984,deutsch_quantum_1991,srednicki_chaos_1994,
  srednicki_approach_1999,rigol_thermalization_2008, DAlessio_fromquantum_2016,borgonovi_quantum_2016} has become the essential framework for reconciling quantum dynamics with statistical mechanics.  In its simplest form, ETH posits that expectation values of local observables in energy eigenstates are smooth functions of the energy eigenvalue in the thermodynamic limit.  
This provides a mechanism for thermalization in isolated quantum systems.
ETH can be understood in terms of random matrix theory: ergodic quantum systems are essentially well described by random matrix ensembles at least where local observables are concerned.
This leads to the above smoothness condition, and also predicts the correct scaling of statistical deviations from
it with system size~\cite{beugeling_finitesize_2014, DAlessio_fromquantum_2016,khaymovich_eigenstate_2019,ikeda_how_2015}.  
In particular, the expectation values of local observables have
gaussian distributions --- the distribution shape is an important characteristic of ETH behavior~\cite{khaymovich_eigenstate_2019}.

It turns out that some disordered interacting systems can avoid thermalization if disorder is strong
enough. Such a nonequilibrium phase of matter is called the Many Body Localized (MBL) phase
\cite{anderson_absence_1958,fleishman_interactions_1980,georgeot_integrability_1998,basko_metalinsulator_2006,gornyi_interacting_2005,oganesyan_localization_2007,znidaric_many-body_2008,berkelbach_conductivity_2010,pal_many-body_2010,schreiber_observation_2015,luitz_many-body_2015,nandkishore_mbl_2015,altman_universal_2015,imbrie_many-body_2016,abanin_recent_2017,alet_many-body_2018,pietracaprina2019hilbert}. 
In this phase, transport is completely halted and the system becomes a perfect insulator. In
particular ETH is not valid~\cite{pal_many-body_2010,luitz_many-body_2015}.  The current theoretical understanding of the
MBL phase relies on the emergence of integrability via a complete set of local integrals of motion
(LIOM)~\cite{serbyn_local_2013,huse_phenomenology_2014,imbrie_many-body_2016,ros_integrals_2015,imbrie_diagonalization_2016}.
For instance, this theory accounts for the failure of thermalization, the area law of entanglement
entropy in infinite temperature eigenstates~\cite{bauer_area_2013} and the logarithmic growth of entanglement entropy after a
quench~\cite{znidaric_many-body_2008,bardarson_unbounded_2012}.

Even though the existence of the MBL phase is by now well established in one dimensional systems, in both theory \cite{basko_metalinsulator_2006,imbrie_diagonalization_2016,imbrie_many-body_2016} and experiments~\cite{schreiber_observation_2015,smith_many-body_2016}, the nature of the localization-delocalization transition remains an active area of research. 
One outstanding question is the universality of anomalous thermalization~\cite{luitz_anomalous_2016,roy_anomalous_2018}, characterized by sub-diffusive  transport, close to the localization transition coming from the ergodic side  \cite{bar_lev_dynamics_2014,agarwal_anomalous_2015, potter_universal_2015,vosk_theory_2015,luitz_extended_2016,lev_transport_2017, bordia_probing_2017,luitz_ergodic_2017, znidaric_interaction_2018, kozarzewski_spin_2018, schulz_energy_2018,lezama_apparent_2019}. 
Moreover, there is evidence that distributions of diagonal matrix elements of the local (globally conserved) density develop heavy tails in this anomalous thermal phase \cite{luitz_long_2016}. It has been suggested that the latter is connected to the sub-diffusive transport and, in addition, could be described by a modified version of ETH~\cite{luitz_anomalous_2016,roy_anomalous_2018}. However it is not clear whether power law tails in the distribution of local operators are a general feature of the sub-diffusive regime.

In this work, we consider the probability distributions of local correlation functions in mid-spectrum energy eigenstates to determine their features in the ergodic as well as in the MBL phase.  
While the gaussian shape of these distributions is a central property of pure ETH, their behavior is equally important to characterize non-ergodic phases, in particular the MBL phase.
Considering the Heisenberg model with random on-site fields, we present and analyze the
distributions for two-point operators: spin-flip and $S_i^z S_{i+r}^z$ operators. 
Due to the U(1) symmetry of the XXZ model, there are no other non-vanishing two-point correlators.
Furthermore, we carry out quantitative quasi-degenerate perturbation-theory calculations (around the limit of infinite
disorder) to explain various features of the distributions in the MBL phase.

The energy eigenstate distributions of spin-flip and $S_i^z S_{i+r}^z$ correlators considered in this paper are gaussian for small disorder strengths but acquire significant weight in the tails already for intermediate disorder $W\approx 2 < W_c\approx 3.7$, $W_c$ being the critical disorder strength to enter the MBL phase. Despite the heavy tails in the thermal regime, the variance of the distribution falls off with increasing system size for fixed $W$ up to the critical value $W_c$. 
Within the MBL regime, $W>W_c$, the variation of the distribution with increasing system size is negligible and the distribution has features not present in the thermal regime. In particular, the spin-flip operator distribution exhibits a sharp peak at zero with smaller satellite peaks on each side and further small peaks at the edge of the distribution, $\pm 1/2$. Perturbation theory captures the form of the large disorder distribution quantitatively.

Perturbative methods to describe localization-delocalization phenomena in condensed matter physics have a long history dating back to Anderson's seminal work and continuing today to address questions relating to MBL \cite{anderson_absence_1958,ros_integrals_2015,scardicchiothiery,basko_metalinsulator_2006,gornyi_interacting_2005,imbrie_many-body_2016}. 
In the context of MBL, two of the main questions were to systematically construct the local integrals of motion that are thought to characterize the MBL phase and to estimate the transition point between MBL and the thermal phase. Both can be achieved by computing perturbative corrections to the mutually commuting occupation numbers at infinite disorder under the constraint that the corrections themselves continue to commute~\cite{ros_integrals_2015}. 
This has to be done to infinite order within some suitable approximation to capture possible delocalization. To make sense of the perturbation theory, as in the case of Anderson localization, there are resonances that lead to naive divergences coming from states close in energy that are mixed by hopping in the latter case and interactions in the former. 
In both cases, the divergences may be resolvable giving the perturbation theory a finite radius of convergence. 
Resolving these divergences amounts to diagonalizing the resonating configurations exactly. 

The perturbation theory discussed in this paper is an expansion in the hopping part of the Hamiltonian around the infinite disorder limit. We carry out the expansion to low orders to be quantitative at large disorder for our finite system and to capture the main qualitative features for smaller disorder within the MBL regime. In the spirit of earlier works, we deal with resonances in the non-degenerate perturbation theory by diagonalizing exactly on the resonant subspaces.

In Section \ref{sec:background} we present the model and the local operators whose correlation functions we study.  
Section~\ref{sec:exp_pm} focuses on the
spin-flip operators across the whole phase diagram: first to nearest neighbor, then the further neighbor spin-flip operator distributions. 
The form of the spin-flip operator distributions in the MBL regime are rationalized within perturbation theory in Section~\ref{sec:exp_pm_pt}.
We then turn to the $\langle S_i^z S_{i+r}^z \rangle$ correlators (Section~\ref{sec:exp_zz}) and the corresponding connected correlators (Section~\ref{sec:exp_zz_connected}), in both cases showing the development of heavy tails at $W\approx 2.0$ and the evolution of these distributions into the MBL regime. 
We highlight the distinctive form of the distributions in the Anderson localized phase and the difference with the corresponding MBL distributions (Section~\ref{sec:exp_zz_ai_mbl}). 
Finally, we compute the distribution using quantitative perturbation theory showing, once again, that it captures well the form of the distributions at strong disorder (Section~\ref{sec:exp_zz_pt}).

\section{Background Material}
\label{sec:background}

\subsection{Model}

\label{sec:model}

We study the canonical XXZ model with random fields $h_i$ along the $z$ direction, 
\begin{eqnarray}
\label{Model}
H = \sum_{i=0}^{L-1}\!\dfrac{J}{2}\!\left(S^{+}_{i}S^{-}_{i+1}\!+S^{-}_{i}S^{+}_{i+1}\right)\!+\Delta S^{z}_{i}S^{z}_{i+1}\!-h_{i}S^{z}_{i}.
\end{eqnarray}
This model -- which is widely studied in the context of MBL~\cite{znidaric_many-body_2008,pal_many-body_2010,berkelbach_conductivity_2010,
	bardarson_unbounded_2012,deluca_ergodicity_2013,serbyn_local_2013,bauer_area_2013,
	nanduri_entanglement_2014,lev_dynamics_2014,
	luitz_many-body_2015,nandkishore_mbl_2015,agarwal_anomalous_2015,lev_absence_2015,bera_many-body_2015,torres-herrera_dynamics_2015,  
	luitz_extended_2016,serbyn_spectral_2016,singh_signatures_2016,
	enss_mbl_2017,tomasi_quantum_2017,bera_density_2017,lezama_one-particle_2017,
	alet_many-body_2018,
	herviou_multiscale_2019,sierant_level_2019,vsuntajs_quantum_2019,serbyn_thouless_2017,maksymov_energy_2019} -- can be mapped to a spinless fermion model with nearest neighbor hopping $J/2$,
interaction term $\Delta$ and on-site potential $h_i$. In this paper, periodic boundary conditions
are set, $J=1$ is fixed throughout the paper and the fields $h_{i}$ are distributed uniformly in $[-W,W]$
with disorder strength $W$. We focus mainly on the isotropic
point $\Delta=1$ (interacting spinless fermions) of the parameter space. However in Section \ref{AI}
we compare also to results for various $\Delta$, including the point $\Delta=0$ (free spinless fermions).  The operators
$S^\alpha_i=\sigma^\alpha_i/2$ are spin 1/2 operators, with $\alpha=0,x,y,z$ and $i$ is the site
index.

The total magnetization $M=\sum_{i=0}^{L-1}S^z_i$ along the $z$ direction is conserved.  We
therefore focus on the largest magnetization sector $M=0$ for even system sizes $L$, corresponding
to the Hilbert space dimension $\text{dim}(\mathcal H)=\text{binom}(L,\lfloor{L/2\rfloor})$. For
each disorder realization $\{h_0,\dots, h_{L-1}\}$, we obtain $\gtrsim 50$ eigenstates closest to
the energy target $(E_\text{max}+E_\text{min})/2$ ($E_\text{min}$ being the ground state energy and
$E_\text{max}$ the highest energy of the sample) using a state-of-the-art shift-invert
code \cite{luitz_many-body_2015,pietracaprina_shift-invert_2018}.  We consider the probability
distributions of various eigenstate expectation values of local operators, i.e. the diagonal matrix
elements of these operators in the eigenbasis of the Hamiltonian.  Our results are histograms over
at least $10^{3}$ disorder realizations for each system size $L$ and disorder strength $W$, we also
calculate the correlators for all sites $i\in[0,L-1]$ to improve the statistics, since the average
over disorder is translation invariant.  The mid-spectrum states of this model are known to exhibit
two dynamical phases \cite{pal_many-body_2010,luitz_many-body_2015}: at low disorder ($W \lesssim
3.7$) they obey the ETH, while at strong disorder ($W \gtrsim 3.7$) all eigenstates are many-body
localized (MBL).

\subsection{Operators}
\label{sec:operators}

In previous works in the context of many-body localization and the MBL transition, the distributions
of local operators were considered, mostly focussing on distributions of diagonal or off-diagonal
matrix elements of simple local observables such as the local magnetization (or number density in
the language of spinless fermions) $\bra{n} S_i^z \ket{n}$, where $\ket{n}$ is a central
eigenstate of the Hamiltonian 
\cite{luitz_long_2016,luitz_anomalous_2016}.  In this work, we consider  more complicated operators
given by two point correlation functions.  First, we consider the correlators
\[
\bra{n} S^+_{i} S^-_{i+r}/2
+h.c. \ket{n} = \bra{n} F_{i,i+r} \ket{n},  
\]
i.e., the matrix elements of spin-flip operators $F_{i,i+r}$. 
We also consider diagonal  two-point correlators, namely $\bra{n} S^z_i S^z_{i+r} \ket{n}$ and its
`connected' version 
\[
\bra{n} S^z_i S^z_{i+r} \ket{n}_c =
\bra{n} S^z_i S^z_{i+r} \ket{n} - \bra{n} S^z_i \ket{n} \bra{n} S^z_{i+r} \ket{n}.
\]

For $r=1$, the first expression above corresponds to the kinetic energy
density, while the second expression is the interaction energy density in the language of spinless
fermions.  The connected correlator $\bra{n} S^z_i S^z_{i+r} \ket{n}_c$ was previously considered in
Ref. \onlinecite{pal_many-body_2010}.

\section{Eigenstate expectation values of $S^+_{i} S^-_{i+r} +h.c. $}\label{FlipFlop}
\label{sec:exp_pm}

\begin{figure}[h]
	\centering
	\includegraphics{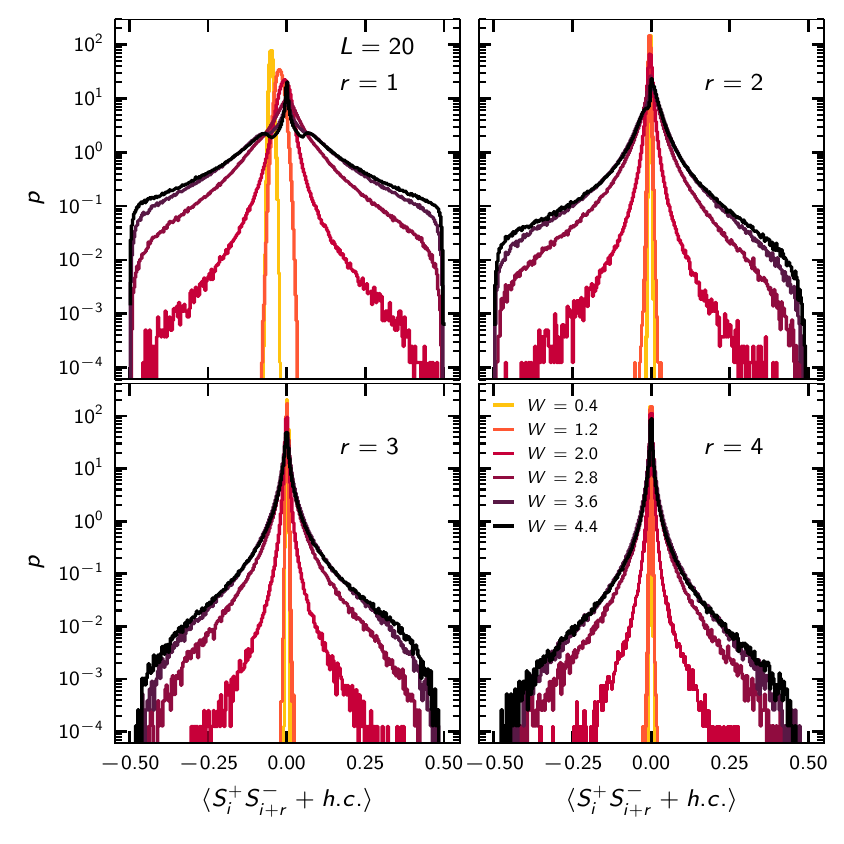}
	\caption{Probability density of eigenstate expectation values $\bra{n} F_{i,i+r} \ket{n}$ for distances $r =1,2,3,4$. The histogram was taken over $\gtrsim 50$ eigenstates, $>1000$ disorder realizations and all positions $i$ in the chain of length $L=20$. In each panel the histograms for the same set of representative disorder strengths $W\in\{0.4, 1.2, 2.0, 2.8, 3.6, 4.4\}$ are shown with the same color code (legend in lower right panel). }
	\label{Distr_SxSy}
\end{figure}

In Fig. \ref{Distr_SxSy} we show the probability distribution of eigenstate expectation values of $ F_{i,i+r} = S^+_{i} S^-_{i+r}/2 +h.c.$ for a system of size $L=20$ and different disorder strengths $W$. 

\subsection{Nearest neighbor flip}
\label{sec:exp_pm_nn}

We start by considering the special case $r=1$, where the operator $ F_{i,i+1}$ corresponds to
the kinetic energy per bond. In the thermal phase at weak disorder, we expect this operator to
follow ETH and be distributed according to a gaussian distribution, which is true to very good
precision.

We observe that at weak disorder ($W\lesssim 2$) and $r=1$, the mean of the distribution is slightly negative, a result of the fact that the eigenstates of the Hamiltonian we consider are in the center of the spectrum, and correspond to high but finite temperature due to the asymmetry of the density of states (cf. Appendix~\ref{sec:e_dependence_loc_op} for an analysis of the energy dependence). States corresponding to strictly infinite temperature correspond to energies given by $\tr  (H_{M=0})/ \text{dim}(\mathcal{H}_{M=0}) = -\frac{L}{4L-4}$, where ${H}_{M}$ is the Hamiltonian matrix in the zero magnetization sector. Such states have a zero mean for traceless operators like $ F_{i,i+r}$. 
Zero mean distributions are recovered at intermediate  disorder where the asymmetry of the spectrum is less pronounced and the energy of the eigenstates we consider is indeed close to $-\frac 1 4$ for large $L$. 

At intermediate disorder $W\approx 2$, we observe the development of heavy tails in the distribution, very similar to the situation for the distribution of $\bra{n} S_i^z \ket{n}$ studied in Ref. \onlinecite{luitz_long_2016}, confirming that the presence of such tails appears to be a generic feature at intermediate disorder in the thermal phase. We note that heavy tails are also observed in the $S_i^zS_{i+r}^z$ correlation function studied in Sec.~\ref{sec:exp_zz}.

At strong disorder $W>W_c$ in the MBL phase, we find a strikingly different distribution of the spin-flip operator expectation values $ F_{i,i+1}$; 
 it features a pronounced central peak at zero, accompanied by two minima adjacent to it, which are framed by two satellite peaks, before the probability density $p(\bra{n}  F_{i,i+1} \ket{n})$ decays towards the edges of its domain $\left[-\frac{1}{2}, \frac {1}{2}\right]$. We have found that this intriguing shape persists at strong disorder and can be explained using perturbation theory, a discussion of which we postpone to the end of this section.

\begin{figure}[h]
	\centering
	\includegraphics{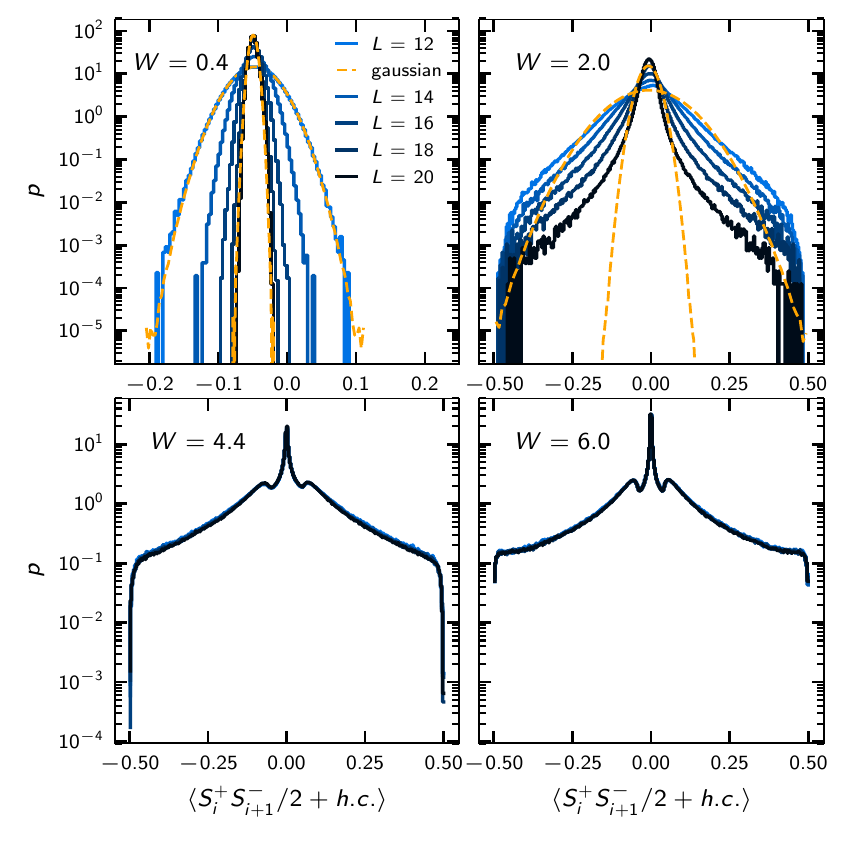}
	\caption{Finite size dependence of the probability density of eigenstate expectation values of the nearest neighbor flip operator $\bra{n} F_{i,i+1} \ket{n}$. As in Fig. \ref{Distr_SxSy}, the histogram is taken over $\gtrsim 50$ eigenstates per disorder realization, $>1500$ disorder realizations and all positions $i$ in the chain. The dashed blue line shows the gaussian distribution computed with the mean and variance belonging to the data $L=12,20$.
		 }
		 
	\label{Distr_SxSy_L}
\end{figure}

\begin{figure}[h]
	\centering
	\includegraphics{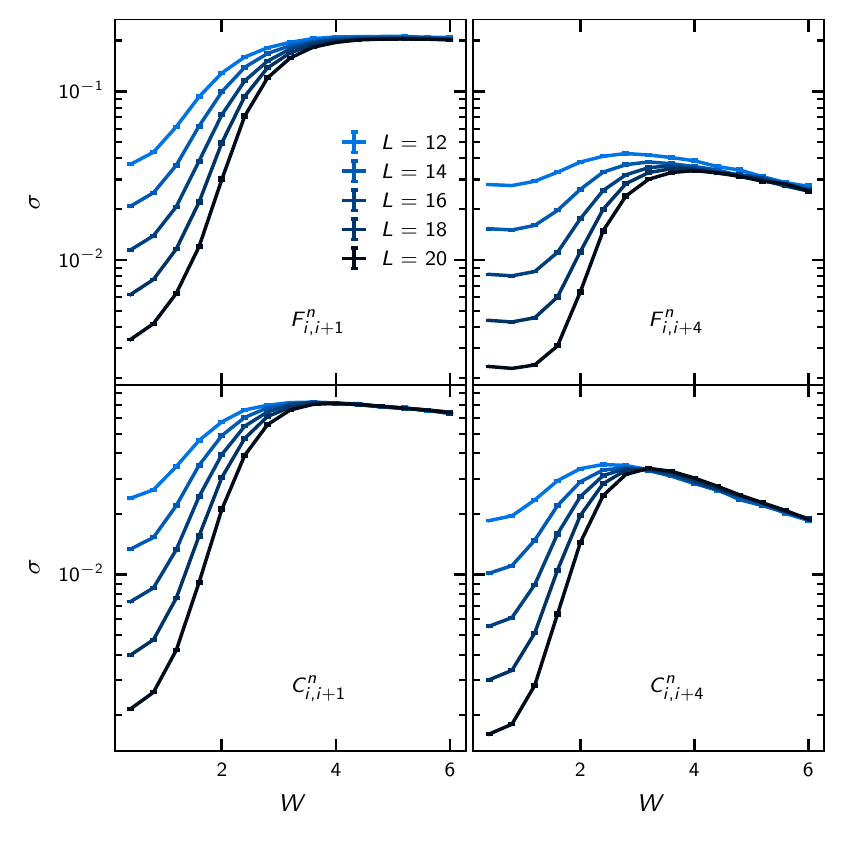}
    \caption{Standard deviation of the difference between adjacent eigenstates of spin-flip operator (upper panels) $F^{n}_{i,i+r}=\bra{n+1}F_{i,i+r}\ket{n+1}-\bra{n}F_{i,i+r}\ket{n}$ and connected $z$ correlation (lower panels) $C^{n}_{i,i+r} = \bra{n+1}S^{z}_{i}S^{z}_{i+r}\ket{n+1}_c - \bra{n}S^{z}_{i}S^{z}_{i+r}\ket{n}_c$, where we have used the shorter notation $ \bra{n}S^{z}_{i}S^{z}_{i+r}\ket{n}_c=\bra{n}S^{z}_{i}S^{z}_{i+r}\ket{n}-\bra{n}S^{z}_{i}\ket{n}\bra{n}S^{z}_{i+r}\ket{n}$.  
   	Instead of the direct variance of the distributions, we consider differences in adjacent eigenstates as in Ref. \onlinecite{luitz_long_2016} to mitigate the slightly different means of distributions at weak disorder due to energy targets depending on the disorder realization (cf. Appendix~\ref{sec:e_dependence_loc_op}).
    }   
	\label{fig:standard}
\end{figure}

In Fig. \ref{Distr_SxSy_L}, we analyze the system size dependence of the probability density of the
nearest neighbor flip operator $ F_{i,i+1}$ over the whole range of disorder strengths, comparing
distributions for sizes $L=12, 14, 16, 18, 20$. At the weakest disorder $W=0.4$, we find gaussian
probability distributions, with the variance decreasing exponentially in system size $L$, as
expected from ETH (cf. Fig.~\ref{fig:standard}). At intermediate disorder $W=2.0$, the distribution
is no longer gaussian, but the variance still decreases exponentially with size. It appears that the
heavy tails, deviating from the gaussian shape, persist even at large system size, following the
same phenomenology observed for the $S_i^z$ operator in Ref.
\onlinecite{luitz_long_2016,luitz_anomalous_2016,roy_anomalous_2018}.

\begin{figure}[h]
	\centering
	\includegraphics{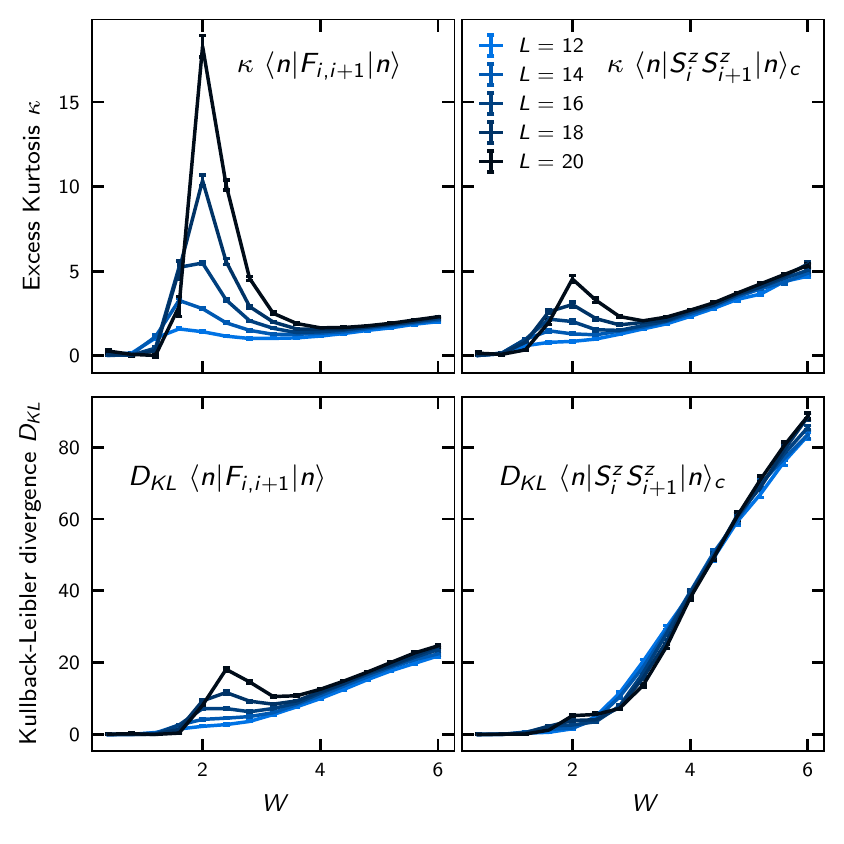}
	\caption{Upper panels: Excess kurtosis $\kappa = (\mu_4 /\sigma^4)-3$ of the distribution of diagonal matrix elements of (left) $\bra{n}F_{i,i+1}\ket{n}$ and (right) connected correlation $\langle  S_i^z  S_{i+1}^z \rangle_c$. A vanishing excess kurtosis corresponds to a gaussian distribution. Lower panels: Kullback-Leibler divergence of the matrix element distributions with respect to a gaussian with same mean and variance. }  
	\label{fig:kurtosis}
\end{figure}

To quantify departures from gaussianity, we compute the excess kurtosis $\kappa = (\mu_4 /\sigma^4)-3$, ($\mu_4$ being the $4$th central moment of the distribution) and the Kullback-Leibler divergence defined by
\begin{equation}
D_{\rm KL} \equiv -\int \D{x}\, P(x) \log\left( \frac{Q(x)}{P(x)} \right)
\end{equation}
where $Q(x)$ is the reference gaussian distribution and $P(x)$ is the computed distribution of the correlator, where the gaussian is defined by the mean and variance of $P(x)$. Results are shown in Fig.~\ref{fig:kurtosis}. Both quantities indicate that the distribution is quantitatively gaussian for $W\lesssim 1.5$ and that they become strikingly less gaussian with a peak at about $W=2$ that increases with system size. Beyond the peak for larger disorder both measures increase smoothly with little system size dependence, indicating strongly nongaussian distributions in the MBL phase.

In the MBL phase at strong disorder, there is no discernible system size dependence of the distribution (Figs.~\ref{Distr_SxSy_L}, lower panels, and \ref{fig:standard}), showing a pronounced maximum at zero, framed by two symmetric satellite peaks, which seem to get closer to each other at stronger disorder.

\subsection{Long distance flip}
\label{sec:exp_pm_fnn}

The flip operator of distant spins $ F_{i,i+r}$, with $r>1$ is not a term of the Hamiltonian and could therefore behave differently. 
We have verified that this is so by examining the energy dependence of the mean of the distribution which is constant over a large range of energies for $r>1$, linear for $r=1$, cf. Appendix~\ref{sec:e_dependence_loc_op}.
For this reason, the mean of the $r>1$ distribution is close to zero at weak disorder. 
At intermediate disorder, the distribution again shows heavy tails, and, in general, the variance decreases with longer distance between the operators, which we attribute to decreasing long distance correlations.
Most interestingly, at strong disorder in the MBL phase and at long distance, the peculiar satellite
peaks of the distribution at $r=1$ disappear, leading to simple, yet heavy tails. Additionally, we see that the standard deviation of both correlation functions at larger distances decreases as function of disorder (Fig.~\ref{fig:standard}) and stay constant at $r=1$. In the limit $W\rightarrow\infty$ the spins are uncorrelated so both standard deviations will go to zero. In this range of disorder, the localization length is big enough for allowing correlations at $r=1$, hence we expect the standard deviation to start decreasing only at large enough disorder.  

The absence of satellite peaks for $r>1$, as well as most of the other features in this and the
preceding subsection, can be understood through perturbation theory in $1/W$, as we describe in the
next subsection.

\subsection{Perturbation theory analysis}
\label{sec:exp_pm_pt}
We have postponed the discussion of the peculiar features of the distribution of the nearest neighbor flip operator $ F_{i,i+r}$ in the MBL phase -- a topic to which we now turn.

\begin{figure}[h]
	\centering
	\includegraphics{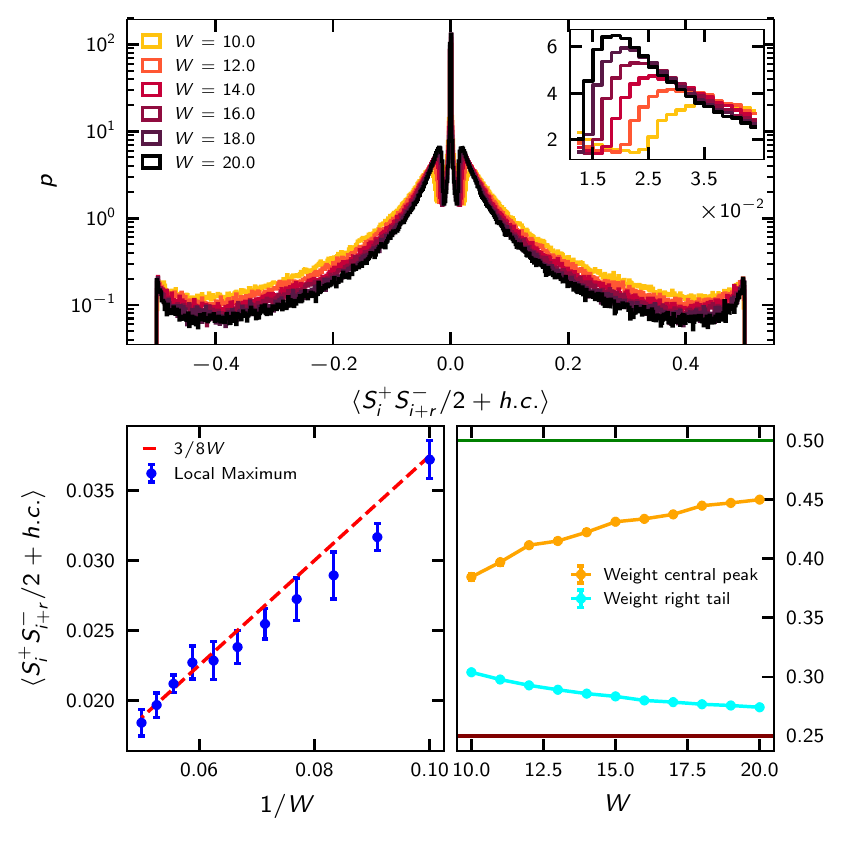}
    \caption{Upper panel: distribution of matrix element $\langle S^{+}_{i}S^{-}_{i+1}/2 +h.c. \rangle$ at large disorder strength $W\geq10$, in the upper right corner the local maximum of the distributions is highlighted. Lower left panel: position of the local maximum as function of the disorder strength. The red dashed line is the exact maximum location extracted from first order perturbation theory and given in Eq. \ref{eq:dist}.  Lower right panel: weight of the distribution at central peak $\int_{-\epsilon}^{\epsilon} \mathrm{d}x \, p(x)$ and weight of the right tail $\int_{\epsilon}^{0.5} \mathrm{d}x\, p(x)$ as function of disorder strength and $\epsilon=0.01$. The weight of the peak at zero tends to $\frac 1 2$ for strong disorder as predicted by perturbation theory.}
	\label{Distr_SxSy_maximum}
\end{figure}

In Fig. \ref{Distr_SxSy_maximum} we have a closer look at its distribution for different (strong) disorder strengths. While the qualitative features (central and satellite peaks) are independent of disorder and apparently characteristic of MBL, there is a quantitative evolution: the satellite peaks become sharper and move towards zero as the disorder $W$ is increased (inset in Fig. \ref{Distr_SxSy_maximum}). Furthermore, at very strong disorder $W>10$, additional peaks at $-\frac 1 2$ and $\frac 1 2$ develop, which are not present at weaker disorder $W\lesssim6$ (cf. Fig. \ref{Distr_SxSy_L}).

As a first step towards a more quantitative analysis, we consider the drift of the position of the satellite peaks as a function of disorder. The lower left panel of Fig. \ref{Distr_SxSy_maximum} shows the estimated peak positions, which are consistent with a $1/W$ dependence, suggesting a perturbative analysis.

At very strong disorder, it is natural to treat the kinetic term of the Hamiltonian as a perturbation of order $1/W$. Noting that the eigenstates of $ H/W$ are equal to those of $ H$, we cast the Hamiltonian in the form

\begin{equation}
 H /W  =  \frac{1}{W} \sum_i S_i^z S_{i+1}^z  + \tilde{h}_i S_i^z   + \frac{1}{W} \sum_i  F_{i,i+1} =  H_0 + \frac{1}{W}  V.
\end{equation}

The scaled fields, $\tilde{h}_i$, are now distributed uniformly in a fixed range $[-1,1]$. The eigenstates of the unperturbed Hamiltonian $ H_0$ are product states and eigenstates of all $ S_i^z$ operators and can therefore be enumerated by their eigenvalues. The eigenenergies of $ H_0$ for each eigenstate can be easily calculated using these quantum numbers.

Naive perturbation theory produces divergences when the spacing between unperturbed energy levels goes to zero. Such divergences -- dubbed resonances -- are unphysical and are resolved by admixing clusters of nearly degenerate states. Resonances are of great importance in disordered systems and become increasingly so as the system size increases. In order to incorporate the effect of resonances from the large disorder limit, we carry out a \emph{mixed degenerate and non-degenerate perturbation theory} for the operator $ F_{i,i+r}$. Details of the perturbation theory are given in Appendix~\ref{sec:pt}. 
In addition to the rather general discussion given in the appendix, we note here various peculiarities of the perturbative calculation of $\bra{\tilde n}  F_{i,i+r} \ket{\tilde n}$ which simplify our task.
In order to obtain a matrix element $\bra{\tilde n}  F_{i,i+r} \ket{\tilde n}$ of an eigenstate $\ket{\tilde n}$ of the perturbed Hamiltonian in perturbation theory, we start with an eigenstate $\ket{n_0}$ of $ H_0$. 
The matrix element $\bra{n_0} F_{i,i+r} \ket{n_0}$ is the zeroth order contribution and is identical to zero because $ F_{i,i+r}$ is off-diagonal in the $z$ basis. 
It therefore contributes to the prominent peak of the distribution of this matrix element at zero. 
More precisely, for $r=1$, states with $\vert\dots 00 \dots\rangle$ or $\vert \dots 11 \dots \rangle$ on the sites $i$ and $i+1$ yield a zero contribution at zeroth and first order in perturbation theory. This accounts for half of the states so we expect the fraction of such states to tend to $1/2$ as $W$ increases and this is indeed what is found (Fig.~\ref{Distr_SxSy_maximum} lower right panel). 
For $r>1$, one must go to higher order in perturbation to obtain any non-vanishing contribution so the central peak is significantly higher. To understand the satellite peaks, we have to go to first order in perturbation theory (cf. e.g. Fig. \ref{Distr_SxSy_maximum}).

\begin{figure}[h]
	\centering
	\includegraphics{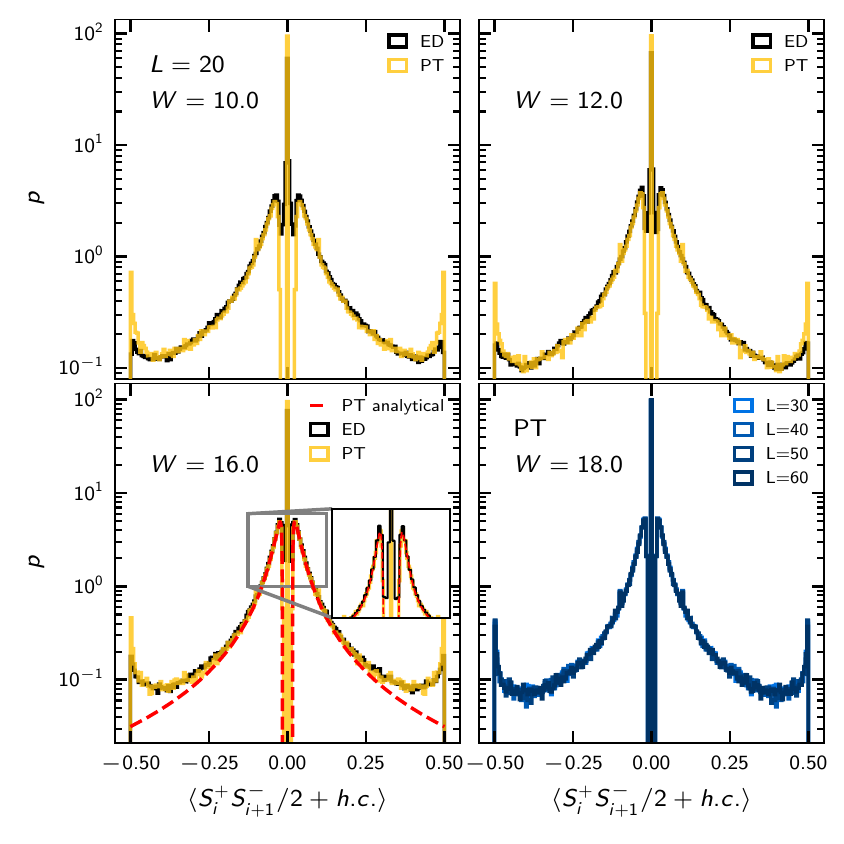}
	\caption{Perturbation theory computation of $\langle S^{+}_{i}S^{-}_{i+1}/2 +h.c. \rangle$ compared the exact result (black). Upper and lower left panels display the results for $L=20$ and $W=10,12,16$. In the inset we can see that perturbation theory distribution is disconnected (see main text). The red dashed line is the analytical form of the probability distribution shown in Eq. \eqref{eq:dist}. Lower right panel: Perturbation theory results for $L=30,40,50,60,70$. }
	\label{PT}
\end{figure}

The eigenstate $\ket{n_0}$ is connected to a set of other eigenstates $\{\ket{k_0} \}$ of $ H_0$ by the perturbation $ V$. By this, we mean that the $\bra{n_0} V\ket{k_0} \neq 0$ for all $\ket{k_0}$ in this set, while matrix elements of $ V$ with all other states vanish. Let us first deal with the case in which all energies $E_{k_0}$ are sufficiently different from $E_{n_0}$, such that in nondegenerate perturbation theory the denominators $1/(E_{k_0}-E_{n_0})$ do not diverge. In this case, we obtain for the matrix element $\bra{\tilde n} F_{i,i+r} \ket{\tilde n}$ up to first order in $1/W$:
\begin{equation}
\begin{split}
	\bra{\tilde n} F_{i,i+r} \ket{\tilde n} &= \frac{1}{W} \sum_{k_0} \bra{k_0} F_{i,i+r}\ket{n_0} \frac{ \bra{n_0}  V \ket{k_0} }{E_{k_0}-E_{n_0}} \\
	 &+ (k_0 \leftrightarrow n_0).
\end{split}
	\label{eq:flipflop_PT}
\end{equation}
From this, we can now understand several features of the distribution: if $\ket{n_0}$ does not have eigenvalues of $ S_j^z$ with opposite sign on sites $j=i$ and $j=i+r$, the matrix element $\bra{\tilde n} F_{i,i+r} \ket{\tilde n}$ vanishes. This implies that due to the incompatibility of $ V$ and $ F_{i,i+r}$ for $r>1$, to first order in $1/W$ all matrix elements vanish, which explains the different behavior of $ F_{i,i+1}$ and $ F_{i,i+r}, r\geq 2$. If the spins on sites $i$, $i+1$ have opposite $S_j^z$ eigenvalues (i.e. they are ``flippable''), there is only one nonvanishing term in the sum, in Eq. \eqref{eq:flipflop_PT}. The matrix element $\bra{n_0}F_{i,i+1}\ket{k_0}=1/2$ in this case giving 
\begin{equation}\label{app5}
\bra{\tilde{n}}F_{i,i+1}\ket{\tilde{n}} =\dfrac{1}{2W\left( E_{n_0}-E_{k_0}\right)}=\dfrac{1}{2W\left( \tilde{h}_{i}-\tilde{h}_{i+1}\right) }.
\end{equation}  
Since the on-site fields have a uniform distribution bounded by $\tilde{h}_i\in [-1,1]$, the expression in Eq. \ref{app5} can be computed. We first note that the lower bound on the matrix element is $1/4W$ as the maximum difference in the fields is $\pm2$. Now, rewriting $x=\bra{\tilde{n}}F_{i,i+1}\ket{\tilde{n}}$ as a random variable that takes values in the range $[-\infty,-1/4W]\cap[1/4W,\infty]$, its probability distribution is:
\begin{equation}\label{eq:dist}
P(x) = \dfrac{4W\vert x\vert -1}{16W^2 \vert x\vert^3}, \hspace{5mm} \vert x\vert\geq 1/4W.
\end{equation}
The maxima of $P(x)$ are located at $x=\pm 3/8W $ 
To summarize, first order nondegenerate perturbation theory explains the presence and weight of the central peak at zero, the presence and location of the satellite peaks at $O(1/W)$, as well as the local minima separating these peaks. The satellite peak stems therefore from the admixture of states which change their energy maximally upon flipping of two neighboring spins.  Fig.~\ref{PT} shows plots of the exact result (black) together with the distribution Eq.~\eqref{eq:dist} (red dashed) showing that the formula captures the exact distribution very well (the delta peak at zero with weight $\frac 1 2$ is not shown in Fig.~\ref{PT}). To examine the agreement in more detail, Fig.~\ref{Distr_SxSy_maximum} shows the close correspondence between the analytical calculation of the satellite peak location and the exact result at least for larger values of disorder. We notice that the perturbation theory produces a higher central peak. This is caused by the missing weight around the central maximum (see inset Fig.~\ref{Distr_SxSy_maximum}) due to the lower bound in magnitude of the matrix elements $\bra{\tilde{n}}F_{i,i+1}\ket{\tilde{n}}$ up to first order (Eq.~\ref{app5}). 

The distribution, Eq.~\eqref{eq:dist}, does not reproduce the small peaks at the edge of the domain of the distribution, close to $\pm \frac{1}{2}$.  To understand the origin of these peaks, we come back to the consideration of the case that the eigenenergy of the state with flipped spins $E_{k_0}$ is close to the energy of $E_{n_0}$, in which case we have a ``resonance'' and nondegenerate perturbation theory breaks down. In this case, we have to use quasi degenerate perturbation theory and include $\ket{n_0}$ and its flipped partner $\ket{k_0}$ in the \emph{model space} of quasi-degenerate states. 
Due to the constraint by the matrix elements of $ F_{i,i+1}$ in the model space, this is the only state which contributes to the model space.
The mixing of these two states leads to the emergence of the peaks at $\pm \frac{1}{2}$ of the distribution. To see this, we consider the zeroth order mixing of quasi-degenerate states through a single spin flip term in the Hamiltonian. This generates pairs of admixed states of the form $\alpha \vert \ldots 10\ldots  \rangle+ \beta \vert \ldots 01\ldots  \rangle$. The flip-flop operator expectation value is ${\rm Re}(\alpha\beta^\star)$.Since the perturbation maximally mixes these quasidegenerate states $\alpha=\pm\beta$ and this accounts for the $\pm 1/2$ peaks.

Our numerical treatment of the exact mixing by quasi-degenerate perturbation theory up to second order in $1/W$ captures also corrections to these features quantitatively and we show the full distributions obtained from it as colored solid histograms in Fig. \ref{PT}. The perturbation theory can be carried out for much larger system sizes than treatable in shift-invert diagonalization and show no visible system size dependence at strong disorder as shown in Fig. \ref{PT}. We conclude that the parts of the distribution of $\bra{n} F_{i,i+1}\ket{n}$ close to zero are due to off-resonant mixing of flippable and not flippable states, while the edges of the distribution close to $\pm \frac 1 2$ reveal the effect of resonances. It should be noted that we do not compute $\bra{\tilde{n}}F_{i,i+r}\ket{\tilde{n}}$ at distances $r>1$ because low order contributions are trivial and higher orders in perturbation theory make the numerical implementation hard to deal with. Considering this, we restrict our perturbation theory computations to operators with $r=1$.

\section{Eigenstate Expectation Values of $S^z_{i} S^z_{i+r}$}
\label{sec:exp_zz}

\subsection{$\left\langle S^z_{i} S^z_{i+r}\right\rangle$ Correlators}

\begin{figure}[h]
    \centering
      \includegraphics{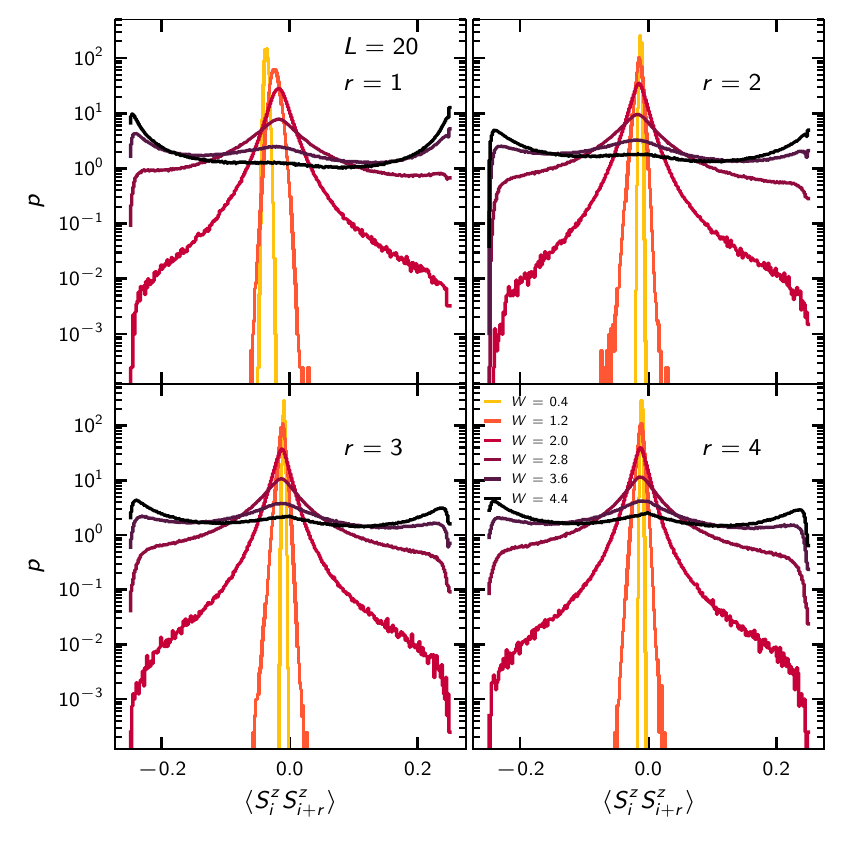}
    \caption{Probability density of eigenstate expectation values $\bra{n} S^{z}_{i}S^{z}_{i+r}  \ket{n}$ for distances $r =1,2,3,4$. The histogram was taken over $\gtrsim 50$ eigenstates, $>1500$ disorder realizations and all positions $i$ in the chain of length $L=20$. In each panel the color corresponds to the disorder strengths as indicated in the legend.}  
    \label{Distr_SzSz}
\end{figure}

We now turn to the $S^z_i$ correlation function.
Fig. \ref{Distr_SzSz} shows the probability distribution of energy eigenstate expectation values of $S^z_{i} S^z_{i+r}$ for a system of size $L=20$, $r=1,2,3,4$ and for different disorder strengths $W$. 
For weak disorder $W\lesssim 1.2$, the distributions are gaussian in accordance with ETH and the
variance of the distribution increases with disorder strength. As remarked in
Section~\ref{sec:exp_pm_nn} and, similarly to the spin-flip correlators studied there, heavy tails
are apparent for disorder strength $W=2$ in the thermal regime. Again, similarly to the spin-flip
correlators, the gaussian mean is displaced from zero and the reason for this displacement is the
same as in that case (cf.\ Appendix~\ref{sec:e_dependence_loc_op}). As one expects in the ETH regime, the variance of the distribution falls off inversely in the Hilbert space dimension (exponential in $L$), which is visible for the case of the connected correlator in Fig. \ref{fig:standard} by the equidistant spacing of the standard deviations for different system sizes on the semilogarithmic scale.

For strong disorder, deep in the MBL regime, the distribution is qualitatively different. 
The central peak is still present but is obscured by a very broad distribution that extends out to the tails where there are more pronounced peaks. 
There are again negligible differences between the distributions for different $L$ within the MBL regime.  
The presence of the outer peaks is simply explained from the strong disorder limit where eigenstates of the Hamiltonian are also eigenstates of the local $S^z_{i}S^z_{i+r}$ operators with eigenvalue $\pm 1/4$. 
The fact that the main new feature of the distribution appears in the large $W$ limit suggests that perturbation theory might be as successful as it was for the spin-flip correlators. We address this question in Section~\ref{sec:exp_zz_pt}.

In order to remove the trivial contribution to the correlation function coming from $\bra{n} S_i^z\ket{n}$ expectation values, we discuss, in the following section, the connected correlation function.

\subsection{Connected Correlators}
\label{sec:exp_zz_connected}

\begin{figure}[h]
	\centering
	\includegraphics{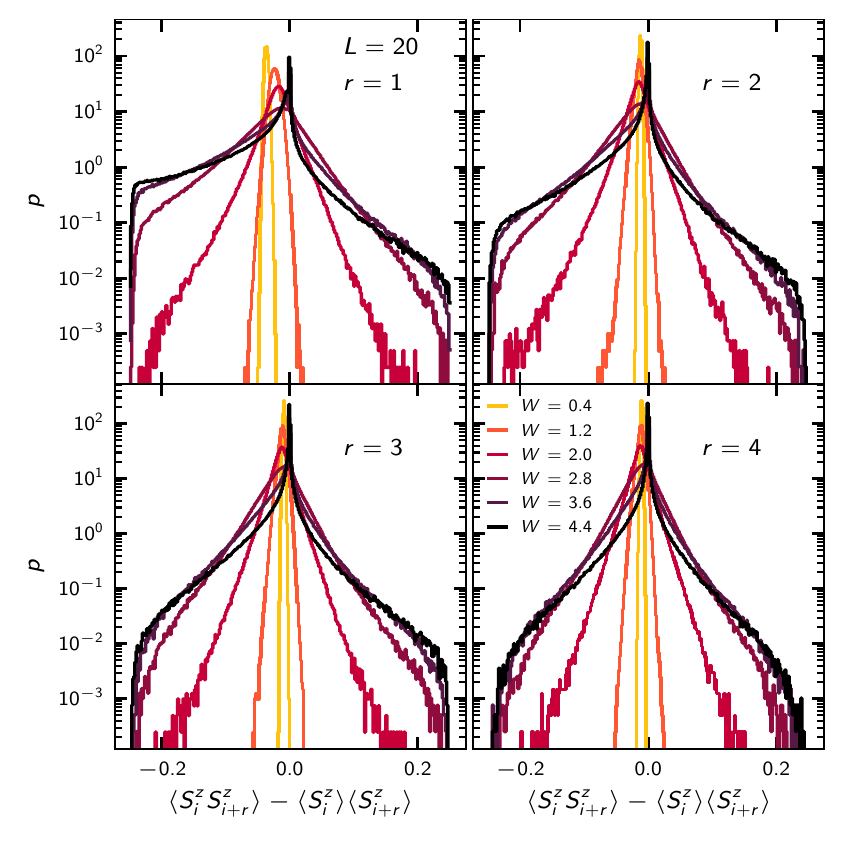}
	\caption{Comparison of the distribution of connected correlators $\langle S^z_iS^z_{i+r} \rangle_c$ in energy eigenstates for different disorder strengths $W$. The histograms include data for different disorder realizations and all lattice sites $i$. For weak disorder ($W\lesssim 1.2$) they display a gaussian distribution.  For strong disorder ($W>3.6$) the distribution exhibits a sharp peak at zero and and heavy tails, biased towards the negative side for short distances $r$.}  
	\label{Distr_SzSz_connected}
\end{figure}

\begin{figure}[h]
	\centering
	\includegraphics{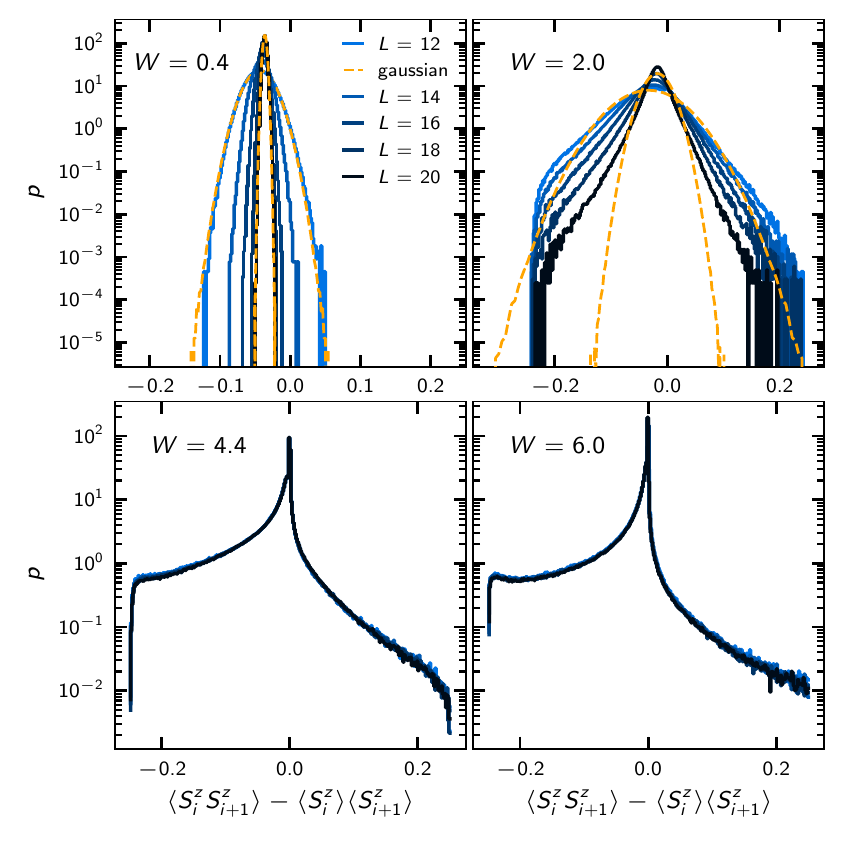}
	\caption{Comparison of the distribution of nearest neighbor connected correlators $\langle S^z_iS^z_{i+1} \rangle_c$ in energy eigenstates for different system sizes. The histograms include data for different disorder realizations and all lattice sites $i$.
	 In the ETH (upper panels) phase the width of the distributions decreases with system size $L$ while in the MBL phase (lower panels) there is no discernible dependence on the system size. The dashed orange lines show gaussian distributions computed with the mean and variance calculated from the data for $L=12,20$. }  
	\label{Distr_SzSz_L}
\end{figure}

Fig.~\ref{Distr_SzSz_connected} shows probability distributions of the connected correlation function for $L=20$ and for different values of $W$ and Fig.~\ref{Distr_SzSz_L} shows distributions for different system sizes. For small $W$, the expectation is that ETH is obeyed and the figures demonstrate that, at least for $W\lesssim 1.2$, the distributions are gaussian (Fig.~\ref{Distr_SzSz_connected}) while the finite size scaling is consistent with random matrix theory (cf. Fig~\ref{fig:standard}). In the ETH regime, there is little variation in the distributions for different $r$ for a given system size -- one merely observes that the mean of the distribution shifts towards zero from $r=1$ to $r>1$ as discussed above due to the different energy dependence of different $r$ operators (cf.\ Appendix~\ref{sec:e_dependence_loc_op}).

For larger values of $W$, a sharp peak forms at zero and persists deep into the MBL phase while the
distributions further depart from gaussianity by acquiring a distinctive asymmetry with higher
weight for negative values of the correlator. For $r=1$, the left-hand-side of the distribution
acquires a shoulder down to $-1/4$ while the positive side tapers off towards $+1/4$. For larger $r$
the shoulder is rounded on the left side, so that the asymmetry is less pronounced. Our analysis in
Appendix~\ref{sec:tails} confirms heavy tails on either side at strong disorder.

In common with other distributions of matrix elements of local operators there is little apparent variation between different system sizes in the MBL regime. In contrast, within the ETH regime for significant values of disorder as exemplified by the $W=2$ data, the central width of the distribution narrows for larger system sizes while weight at the tails remains.

\subsection{Anderson Insulator vs MBL}\label{AI}
\label{sec:exp_zz_ai_mbl}

\begin{figure}[h]
	\centering
	\includegraphics{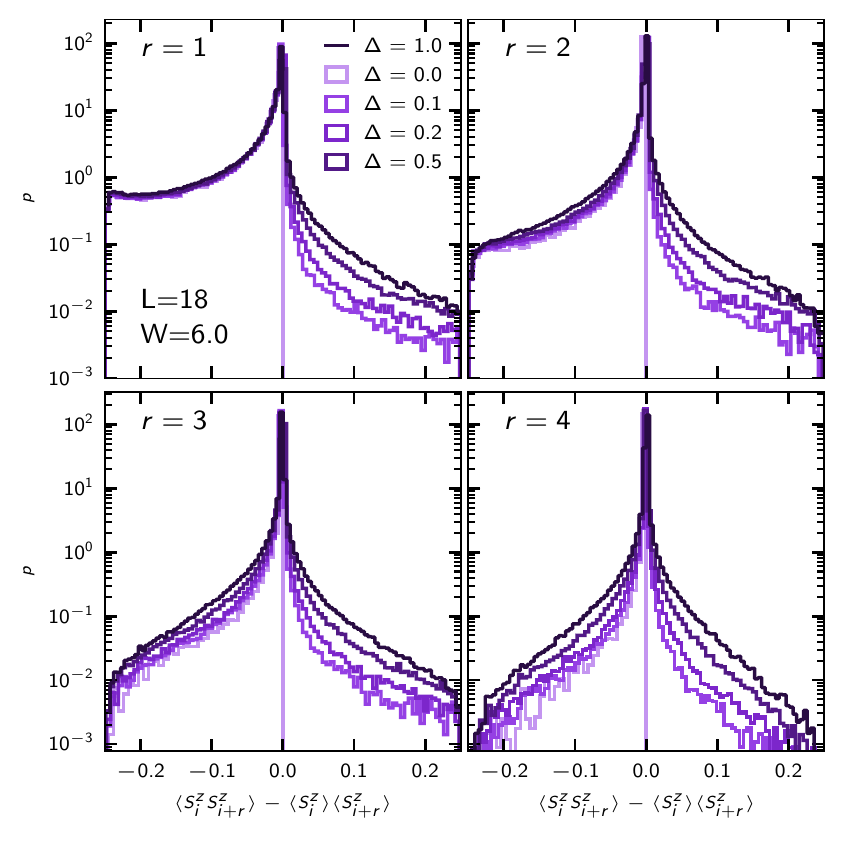}
	\caption{Distribution of the connected correlator $\langle S_i^z S_{i+r}^z \rangle_c$ for the disordered field Heisenberg chain with interaction ($\Delta\neq 0$, MBL) and without ($\Delta=0$, Anderson insulator) at distances $r=1,2,3,4$ and strong disorder $W=6.0$. Remarkably, in the Anderson insulating phase there is no positive weight on the distributions. 
        }
	\label{AI_correlators}
\end{figure}
  
To understand the asymmetry of the distribution of the connected correlator $\langle S_i^z S_{i+r}^z \rangle_c$ for small distances $r$, it is useful to compare to the noninteracting limit.
In Eq.~\ref{Model}, $\Delta=0$ corresponds to the case of an Anderson insulator of noninteracting spinless fermions.
Fig.~\ref{AI_correlators} shows the connected $S_i^z S_{i+r}^z$ correlator for  $\Delta=0.0,0.1,0.2,0.5,1.0$ and for $W=6$ and $r=1,2,3,4$. We see that for distance $r=1$ the distribution of negative correlations has little sensitivity to the value of $\Delta$ while the weight of positive correlations is exactly zero for the Anderson insulator, giving a clear signature where MBL differs from the non-interacting case albeit one where the asymmetry between positive and negative weights persists to $\Delta=1$.

We can understand the vanishing positive weight for the Anderson insulating case for arbitrary $r$
through a straightforward application of Wick's theorem since the Anderson case $\Delta=0$ is a free
spinless fermion model with a gaussian action.  In fermionic language, the $\hat{S}^z_i
\hat{S}^z_{i+r}$ operator takes the form $\left(c_i^\dagger c_i -
\frac{1}{2}\right)\left(c_{i+r}^\dagger c_{i+r} -\frac{1}{2}\right)$.  Using Wick's theorem for
$\Delta=0$, we obtain in any eigenstate of the Hamiltonian: $\langle c_i^\dagger c_i c_j^\dagger c_j
\rangle= \langle c_i^\dagger c_i \rangle \langle c_i^\dagger c_i \rangle - \langle c_i^\dagger c_j
\rangle \langle c_j^\dagger c_i \rangle$.  It follows that the connected correlator is $\langle
n\vert\hat{S}^z_i \hat{S}^z_{i+r}\vert n\rangle_c = -\left\vert \langle n\vert c_i^\dagger
c_{i+r}\vert n\rangle \right\vert^2$, which is necessarily $\leq0$.  This leads to the extreme asymmetry of the connected
correlator distribution at $\Delta=0$.

We see in  Fig.~\ref{AI_correlators} that the asymmetry decreases as $r$ increases.  This can be
understood perturbatively, as discussed in the next section.

\subsection{Perturbation Theory}
\label{sec:exp_zz_pt}

\begin{figure}[h]
	\centering
	\includegraphics{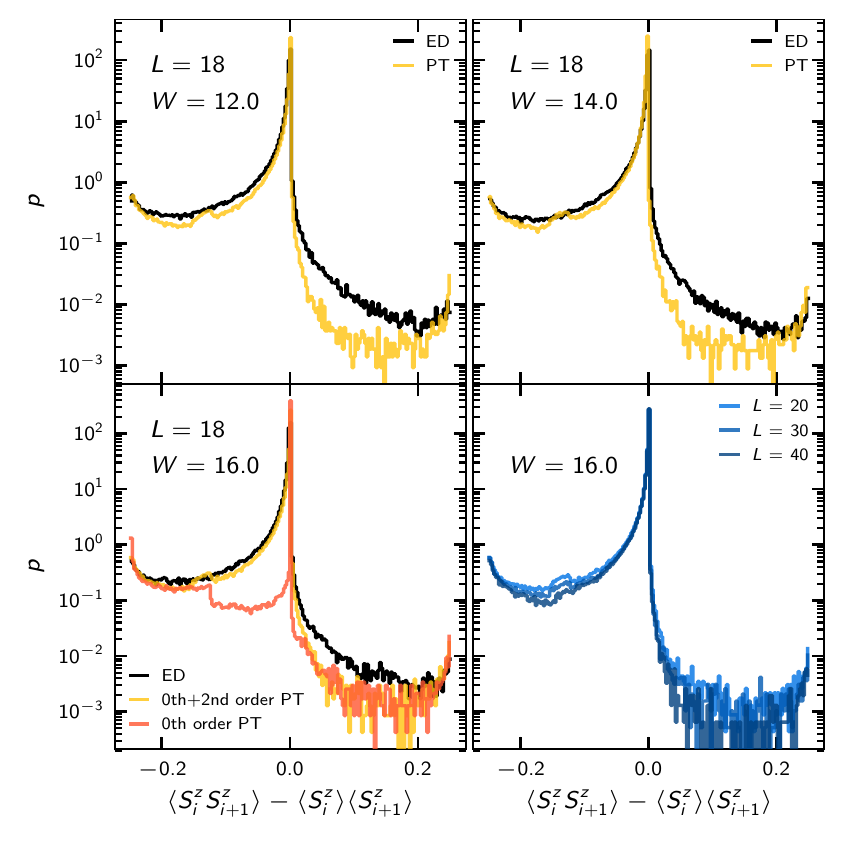}
	\caption{Comparison of the exact $L=18$ distribution (ED) of the connected  $S^z_iS^z_{i+1}$ correlator at strong disorder to the results from quasi-degenerate perturbation theory up to second order in $1/W$. All panels show the results in perturbation theory up to second order, except for the yellow curve in the lower left panel, which shows also only zeroth order degenerate perturbation theory results (mixing within the model space), which appears insufficient to reproduce the full form of the distribution. Lower right: Perturbation theory distributions of the connected correlator for larger system sizes.}  
	\label{PTConnectedSZSZ}
\end{figure}

As in the case of the spin-flip correlator, the distributions of the $S^z$ correlator are well
reproduced at large $W$ by the perturbation theory discussed in Sec.~\ref{sec:exp_pm_pt} and
Appendix \ref{sec:pt}.  This is demonstrated in Fig.~\ref{PTConnectedSZSZ} for the distributions of
connected correlators with $r=1$.  Perturbation theory also provides qualitative physical
explanations for the features reported in the last three subsections, as we elaborate below.

In the infinite disorder limit $W\to\infty$, the correlators $\langle S^z_iS^z_{i+r}\rangle$ and $\langle S^z_i\rangle$ are respectively $\pm 1/4$ and $\pm 1/2$ and the connected correlator simply vanishes, contributing to the sharp peak of the distribution at zero.
Indeed, the numerical data in Fig. \ref{Distr_SzSz}, for large values of $W$, shows peaks at the extreme values of the matrix elements. Similarly the connected correlator distribution in Fig. \ref{Distr_SzSz_connected} exhibits a smooth decay of the central peak at zero towards finite values of the connected correlator.
 
Inspecting the general expressions for matrix elements in mixed degenerate and nondegenerate perturbation theory presented in appendix \ref{sec:pt}, we observe that the zeroth order term coming from mixing within the model space should account for some degree of broadening of the peaks.
The first order terms in $1/W$ trivially vanish, because the $S_i^zS_{i+r}^z$ operator does not connect states in the model space to those outside it, and this is why we proceed to compute the second order contribution. 
Fig.~\ref{PTConnectedSZSZ} shows that perturbation theory to second order (yellow) compares very well to the exact finite size results for $L=18$ and large disorder $W=12, \dots 16$ whereas the zeroth order results (orange) fail to capture the smooth falloff of the distribution to the left of the central peak. We also show results for larger system sizes in Fig.~\ref{PTConnectedSZSZ}, which are not reachable otherwise and show that at large disorder the distributions are essentially converged.

We show below that, while zeroth order (i.e., quasi-degenerate) perturbation theory does not fully account
for the distribution shape of the connected correlators, it does capture the asymmetry.

Zeroth order perturbation theory mixes quasi-degenerate eigenstates of ${H}_0$ connected by ${V} = \sum_i  F_{i,i+1}$ -- in other words states connected by flippable spins $\vert\ldots 01 \ldots\rangle $ or $\vert\ldots 10 \ldots\rangle $ where the ellipses denote some spin configurations. This means, starting from an eigenstate $\ket{n_0}$ of $ H_0$, we can expect mixing with the states $\{  F_{i,i+1} \ket{n_0} \}$, which are quasi-degenerate~\footnote{In practice, we set a cutoff value for the energy difference which we still treat as quasi degenerate of the order of $1/W$. We checked that choosing $2/W$ or larger values does not significantly change the results} with $\ket{n_0}$. 
This means that the model space is then spanned by 
\begin{equation}
\mathrm{D} = \text{span}\left(\ket{n_0}, \{ \ket{i_0} : \ket{i_0} = F_{i,i+1}\ket{n_0} \text{and} \, E_{i_0}\approx E_{n_0} \}\right).
\end{equation}

Let us now try to understand why the distribution of the connected correlator is asymmetric, starting from the case $r=1$.
For simplicity, we consider the case of a two dimensional model space, yielding a state (with $|b|^2 = 1-|a|^2$) of the form:
\begin{equation}
\ket{\psi} = a\ket{\dots \sigma_{i-1}\sigma_{i}\sigma_{i+1} \sigma_{i+2} \dots}+b\ket{\dots \tau_{i-1}\tau_{i}\tau_{i+1} \tau_{i+2}\dots}.
\end{equation}
The connected correlator  is then given by
\begin{equation}
\begin{split}
&4\langle  S_i^z  S_{i+1}^z \rangle_c = |a|^{2} \sigma_{i} \sigma_{i+1} - \left(|a|^{2} - 1\right) \tau_{i} \tau_{i+1}  \\  &- \left[|a|^{2} \sigma_{i} - \left(|a|^{2} - 1\right)\tau_{i} \right] \left[|a|^{2} \sigma_{i+1} -  \left(|a|^{2} - 1\right)\tau_{i+1}\right].
\end{split}
\end{equation}
Inspecting this expression shows that most combinations of spin configurations
$\sigma_i$,$\sigma_{i+1}$,$\tau_{i}$,$\tau_{i+1}$ yield vanishing connected correlators and these
contribute to the central peak. The spin configurations on $i$, $i+1$ that yield non-vanishing
contributions are:
\begin{multline*}
  \sigma_i\sigma_{i+1}\tau_i\tau_{i+1} \in \{ 0011, 1100 \}:
  \\
  4\langle  S_i^z  S_{i+1}^z \rangle_c =
1-(2|a|^2-1)^2 >0.
\end{multline*}
\begin{multline*}
  \sigma_i\sigma_{i+1}\tau_i\tau_{i+1} \in \{ 0110, 1001 \}:
  \\
  4\langle  S_i^z  S_{i+1}^z \rangle_c = (2|a|^2-1)^2-1 <0.
\end{multline*}
This means we obtain two cases for positive correlators and two for negative correlators. 
Evidently the case with a \emph{flippable} pair $\sigma_i\neq \sigma_{i+1}$ (yielding a negative correlator) appears at first order in $V$, since the two states are directly connected through $V$ and are included in the model space if they are quasi-degenerate, \emph{independent of the state of the neighboring spins} $\sigma_{i-1}$ and $\sigma_{i+2}$.
The case of an \emph{aligned} pair $\sigma_i=\sigma_{i+1}$ (yielding a positive correlator) is connected to its flipped partner state $\tau_i=-\sigma_i$ and $\tau_{i+1}=-\sigma_{i+1}$ only in second order of $V$, including a \emph{constraint on the neighboring spins} $\sigma_{i-1}=-\sigma_i$ and $\sigma_{i+2} = -\sigma_{i+1}$. We note that in addition, an intermediate state with one spin flip has to be quasi-degenerate, which is an additional constraint. For simplicity we have left this state out of the discussion.

From these arguments, we conclude that the case of admixed states which yield negative correlations is much more probable than the case yielding positive correlations due to their appearance at different orders in $V$ and, additionally, owing to constraints which reduce the number of possibilities giving obtaining positive correlations.

We now understand that for the case $r=1$, negative correlations are more probable than positive ones to zeroth order in degenerate perturbation theory for two reasons: negative correlations need only one order in $V$, while the appearance of positive weight requires two applications of $V$ and only a specific set of spin configurations can lead to positive correlations, thus reducing their likelihood.

The same set of arguments can now be generalized to the case $r=2$. We see that in this case, we always need to apply $V$ twice to get nonzero (both positive and negative) correlations, however there are more possibilities of having a flippable pair $i,i+2$ (necessary for negative correlations), compared to the possibilities of getting a mixture of $\ket{\dots 0\text{\textsf{x}}0\dots}$ and $\ket{\dots 1\text{\textsf{x}}1\dots}$ (necessary for positive correlations), since in this case the flippability depends on the state \textsf{x} of the middle spin $i+1$. Therefore, also in the case $r=2$, the distribution of the connected correlator is skewed towards negative correlations. In the case of longer distances $r>2$, these constraints become increasingly weak (while requiring an order of $V^r$ to get nonzero correlators), leading to more and more symmetric distributions.

\section{Conclusions}

We have presented the exact energy eigenstate distributions of spin-flip and $S_i^z S_{i+r}^z$ correlators in the disordered XXZ chain across the many-body localization transition.
While -- at very weak disorder -- we find gaussian distributions to very high precision, the distributions depart from gaussianity at intermediate disorder -- still well inside the thermal regime -- through the appearance of heavy tails that persist into the MBL regime.
The presence of these tails correlates to the appearance of sub-diffusive behavior in transport
properties observed in previous studies \cite{luitz_long_2016,luitz_anomalous_2016,roy_anomalous_2018}. 
In the entire thermal regime, the variance of the correlator distributions falls off with increasing Hilbert space dimension as one should expect for operators obeying ETH but significant weight remains in the tails of the distribution and measures of departures from gaussianity including the Kullback-Leibler divergence and the kurtosis show a peak for $W<W_c$ that sharpens with system size. The system size dependence of local operator distributions is negligible inside the MBL regime where ETH fails.

For large disorder, we have carefully investigated the distinctive forms of the correlator
distributions, unraveling various features of the distributions.  We find that strong disorder
perturbation theory can reproduce the full distributions in the MBL phase. We note that our semianalytical perturbation theory scheme should be applicable to other models and could provide information about the effect of resonances in different systems.

For the \emph{spin-flip correlator}, the distributions are highly structured with a central peak at zero, a pair of neighboring satellite peaks with disorder strength dependent positions at $\pm3/8W$ and further maxima at the edge of the distribution at $\pm 1/2$. 
All these features are perfectly captured by a \emph{quantitative strong disorder perturbation theory} that also gives insight into their origins. In particular, (i) the central peak comes from eigenstates where the eigenstate carries no pairs of spins that are flippable by the spin-flip operator, this accounts for $1/2$ of all states at strong disorder 
(ii) the satellite peaks at $\pm 3/8W$ arise from flipped pairs of spins that are maximally pinned by the random field and therefore maximally off resonant 
(iii)  the $\pm 1/2$ peaks \emph{arise from resonances} - strongly admixed quasi-degenerate states. These extremal peaks can only be captured by quasi-degenerate perturbation theory.
Overall, mixed quasi-degenerate and degenerate perturbation theory unifies all contributions and yields an unbiased result, matching the full exact distribution almost perfectly.

The $S_i^z S_j^z$ correlator distribution is more complicated to analyze since we have to go to second order in $1/W$ in our perturbative treatment. Our analysis reveals that for short distances $r=|i-j|$, the correlator is predominantly negative in the MBL phase, since \emph{eigenstates are biased to contain mixtures of flippable neighboring pairs}.
This leads to distributions skewed towards negative weights, most strongly so for noninteracting Anderson Insulators, where no weight for positive correlators is present due to Wick's theorem. Therefore, the $S_i^z S_j^z$ correlator distribution reveals a strikingly different behavior generated by interactions in the MBL case compared to the noninteracting model.

\begin{acknowledgments}
We would like to thank Jeff Rau for useful discussions.
We furthermore acknowledge PRACE for awarding access to 
HLRS’s Hazel Hen computer based in Stuttgart, Germany under grant number 2016153659.
Our code is based on the PETSC\cite{petsc-web-page,petsc-efficient,petsc-user-ref} and SLEPc\cite{hernandez_slepc:_2005} libraries and uses MUMPS\cite{MUMPS1,MUMPS2}.
\end{acknowledgments}


\bibliography{matrixelements}


\clearpage
\appendix
\section{Perturbation Theory}\label{sec:pt}

In this section, we give details of the Rayleigh-Schr\"{o}dinger perturbation theory that is used to obtain the leading contributions in $1/W$ from the transverse exchange term $V$ of the Hamiltonian to the energy eigenstate expectation values of $S_i^z S_{i+r}^z$ and spin-flip correlators (see main text).
We intend for this appendix and the accompanying part of the main text to be self-contained on the method but we refer the interested reader to Lindgren \cite{lindgren_rayleigh-schrodinger_1974} for further details. 

The Hamiltonian from Eq. \eqref{Model} is rewritten $H/W=H_0 + (1/W) V$ where
\begin{align}
H_0 & = \sum_{i=0}^{L-1}(1/W) S^z_{i}S^z_{i+1}-\tilde{h}_{i}S^{z}_{i} \\
V & = \frac{1}{2}\sum_{i=0}^{L-1}\left( S^+_{i}S^-_{i+1} + {\rm h.c.} \right)
\end{align}
and $\tilde{h}_i \in [-1,1]$ are normalized normal fields.

We wish to compute matrix elements of operator ${O}$ in the eigenstate basis where the eigenstates
are computed to some order in perturbation theory in $1/W$.  The eigenstates of interest here are
taken to be highly excited states from the middle of the spectrum.  We start from the eigenstates
$\ket{n}$ of $H_0$, which are product states in the $S^z$ basis, their eigenenergies $E_0^n$ is
trivially obtained from $H_0$.  To compute the corrections to the eigenvectors $\ket{n}$
perturbatively in $1/W$ we should bear in mind that the energies of the product states are
essentially randomly distributed and will mix strongly under the perturbation $V$ if the states are
quasidegenerate.  Loosely speaking, we would like to organize the perturbation theory so that states
coupled by the pairwise spinflips in the perturbation $V$ that are separated by an energy $\Delta
E>1/W$ are treated via non-degenerate perturbation theory while clusters of quasi-degenerate states
(with mutual energy differences $\lesssim1/W$) are treated via degenerate perturbation theory. We
therefore use the formalism of a mixed quasi-degenerate and nondegenerate perturbation theory as
described in Ref. \onlinecite{lindgren_rayleigh-schrodinger_1974}.

We discuss how to organize this calculation up to order $1/W^2$, keeping in mind our main goal: the computation of energy eigenstate expectation values of local operators.

Let us focus on a single $H_0$ eigenstate $\ket{n}$ in one disorder realization that has eigenvalue $E_0^n$. We compute all other states $\{\ket{m}\}$ connected to it by $V$ and $V^2$, i.e. $\bra{n}V\ket{m}\neq 0$ or $\bra{n}V^2\ket{m}\neq 0$. 
Now, we define the {\it model space}, D, to consist of all states in this set ${\ket{m}}$ (including the ``parent state'' $\ket{n}$), and with $\vert E_k - E_n\vert < \alpha/W$. We have experimented with the threshold $\alpha$ to include states in the model space and the results presented in the paper are essentially identical for $\alpha\in[1,3]$.
Let $\mathbb{P}$ be the projector onto the model space D and $\mathbb{Q}=1-\mathbb{P}$ the projector on its complement $\overline{\text{D}}$.

Suppose $\ket{\Psi_n}$ is an exact eigenstate of the full Hamiltonian that has a nonvanishing
projector into the model space $\ket{\Psi^D_\lambda}=\mathbb{P}\ket{\Psi_\lambda}$.  We introduce the inverse wave operator $\Omega = \Omega^{(0)}+\Omega^{(1)}+\Omega^{(2)}+\ldots$ that is expanded in powers of $V$. 
The action of $\Omega$ is to rotate a state in the model space into the full space eigenfunction $\ket{\Psi_\lambda}=\Omega\ket{\Psi^0_\lambda}$. So now if we compute an eigenstate $\ket{v^D_{\lambda}}$ of the effective Hamiltonian
\begin{equation}
H_{\rm eff} \equiv \mathbb{P}H\Omega = H_{\text{eff}}^{(0)} +H_{\text{eff}}^{(1)} + H_{\text{eff}}^{(2)} + \dots
\end{equation}
this is nothing but the projection of the exact wavefunction onto the model space. The eigenfunctions of $H_{\rm eff}$ are generally non-orthogonal and the effective Hamiltonian non-Hermitian (albeit with real eigenvalues). To obtain the eigenstates of $H$ we lift them out of the model space by acting with $\Omega$: $\ket{\Psi_{\lambda}}=\Omega\ket{v^D_{\lambda}}$. Note that $\ket{v_\lambda^D}\in \text{D}$ is a vector of dimension $\text{dim}(\text D) \ll \text{dim}(\mathcal H)$, whereas $\ket{\Psi_\lambda}\in \mathcal{H}$ is a vector in the entire Hilbert space $\mathcal H$.

The problem is now to compute $\Omega^{(n)}$. One may show that \cite{lindgren_rayleigh-schrodinger_1974}
\begin{align}
\Omega^{(0)} & = \mathbb{P} \\
\Omega^{(1)} & = S\left( V \mathbb{P} \right) \\
\Omega^{(2)} & = S\left( V \Omega^{(1)} \right) - S\left(\Omega^{(1)}  V\mathbb{P}  \right)
\end{align}
where $S$ is defined by
\begin{equation}
\langle k \vert SA\ket{m} \equiv \frac{\langle k \vert A\ket{m}}{E_0^m - E_0^k}.
\end{equation}

We now spell out the procedure for computing eigenstate matrix elements of local operators to first and second order in perturbation theory for mid-spectrum states deep in the MBL phase. 

The required steps are the following:
\begin{enumerate}
	\item Select a random ``parent state'' $\ket{n}$, which is an eigenstate of the unperturbed Hamiltonian $H_0$. Its energy $E_0^n$ will typically lie in the middle of the spectrum of $H_0$.
	\item Generate the ``family'' of states $\ket{m}$ connected to $\ket{n}$ by the perturbation $V$ (i.e. by neighboring pairwise spin flips).
	\item Compare the energies of all states $\ket{m}$ with the parent state $\ket{n}$, include  $\ket{n}$ and all states $\ket{m}$ which are quasidegenerate (energy difference $<\alpha/W$) with $\ket{n}$ in the model space D.
	\item For each state $\ket{m}$ which is added to D, its family has to be created and energy differences have to be checked again, possibly including more states in the model space D. States which are well separated from the model space states are included in the complement $\overline{\text{D}}$.
	\item Once this iterative process stops (for the spin-flip operator expectation values further constraints can be used, which simplify this), the effective Hamiltonian $H_\text{eff} \in \mathbb{C}^{\text{dim}(\text{D})\times\text{dim}(\text{D})}$ is calculated (see below).
	\item Next the eigenstates of $H_\text{eff}$ are computed and we pick one eigenstate $\ket{v_\lambda^\text{D}}$ at random.
	\item In the next step, the eigenstate of $H_\text{eff}$ is promoted to the full Hilbert space, using the perturbation expansion of the wave operator $\Omega$ up to the required order. This step can be skipped using the expressions derived below to directly obtain the eigenstate expectation values of the operators we are after.
\end{enumerate}
We note that since we are dealing with central eigenstates of a quantum spin chain with a dense spectrum, there is no separation of clusters of eigenvalues of $H_0$ from other parts of the spectrum. Therefore in principle the procedure above is not guaranteed to yield a model space which is of smaller dimension than the full Hilbert space. It turns out, however, that for the spin-flip correlator, additional selection rules (i.e. the removal of terms yielding zero contributions) guarantee a small model space. For the $S_i^z S_{i+r}^z$ correlators this is not true, and we typically find much larger model spaces, which are however still significantly smaller than the full Hilbert space.

Our procedure is designed to yield the minimal set of states necessary to get the non-vanishing contributions to matrix elements to a given order in perturbation theory.

\subsection{Zeroth Order Perturbation Theory}

The zeroth order effective Hamiltonian is nothing but the projection $H_\text{eff}^{(0)} = \mathbb{P}H\mathbb{P}$ into the model space.
Its matrix elements are therefore $\ket{m},\ket{m'}\in \text{D}$:
\begin{equation}
\bra{m}H_{\text{eff}}^{(0)} \ket{m'} = \bra{m}H \ket{m'}.
\end{equation}
An eigenstate $\ket{v_\lambda^D}$ of $H_{\text{eff}}^{(0)}$ with eigenvalue $\lambda$ then yields the corresponding eigenstate in the full Hilbert space
\begin{equation}
\ket{\Psi_\lambda} = \sum_{\ket{m}\in \text{D}} \braket{m|v_\lambda^\text{D}} \ket{m} = \sum_{m\in \text{D}} v_m^\lambda \ket{m}.
\end{equation}
From this, we may now calculate the zeroth order contribution to the expectation value of an operator $O$:
\begin{equation}
\bra{\Psi_\lambda} O \ket{\Psi_\lambda} = \sum_{m,m' \in \text{D}} v_{m'}^{\lambda,*} v_{m}^\lambda \bra{m'} O \ket{m}.
\end{equation}

\subsection{First Order Perturbation Theory}

The first order effective Hamiltonian is constructed as follows 
\begin{equation}
\langle m \vert H_{\rm eff}^{(1)} \vert m'\rangle =  \sum_{k\in\overline{\text{D}}} \frac{\langle m \vert H \vert k\rangle \langle k \vert V  \vert m'\rangle }{E_0^{m'} - E_0^k} + \ldots
\label{eq:Heff1}
\end{equation}
where ${\ket{k}}$ lives in the complement  space $\overline{\mathrm{D}}$ and $\ket{m}$, $\ket{m'}$
live in the model space $\mathrm{D}$.

To first order, the total effective Hamiltonian is then given by
\begin{equation}
H_\text{eff} = H_\text{eff}^{(0)} + H_\text{eff}^{(1)}\quad  \in \mathbb{C}^{\text{dim}
(\text{D}) \times \text{dim}(\text{D}) }.
\end{equation}

The eigenstates $\ket{v_\lambda}=\sum_{\mathrm{D}} v^{\lambda}_m \ket{m}$ of $H_\text{eff}$ with eigenvalue $\lambda$ can now be lifted into the full space by acting with $\Omega$ to first order: $(\Omega^{(0)}+\Omega^{(1)})\ket{v_\lambda}$. Thus:
\begin{equation}
\ket{\Psi_\lambda} = \ket{v_\lambda} + \sum_{\tiny\begin{array}{c} k\in \overline{\mathrm{D}} \\  m\in \mathrm{D} \end{array}} v^{\lambda}_m \frac{\langle k \vert V \vert m\rangle   }{E_0^m - E_0^k} \ket{k}.
\end{equation}

Then the diagonal matrix elements of operator ${O}$ are
\begin{align}
\langle \Psi_\lambda \vert {O} \ket{\Psi_\lambda} = \sum_{m,m' \in\mathrm{D}} v_m^{\lambda} v_{m'}^{\lambda \hspace{0.5mm}*} \langle m' \vert {O} \vert m \rangle \nonumber \\
+  \sum_{\tiny\begin{array}{c} k\in \overline{\mathrm{D}} \\  m,m' \in \mathrm{D} \end{array}} v_m^{\lambda} v_{m'}^{\lambda \hspace{0.5mm}*} \langle m' \vert {O} \vert k \rangle \frac{ \langle k \vert V\vert m\rangle}{E_0^m - E_0^k} \nonumber \\
+  \sum_{\tiny\begin{array}{c} k\in \overline{\mathrm{D}} \\  m,m' \in \mathrm{D} \end{array}} v_m^{\lambda} v_{m'}^{\lambda \hspace{0.5mm}*} \langle k \vert {O} \vert m \rangle \frac{ \langle m' \vert V\vert k\rangle}{E_0^{m'} - E_0^k}
\end{align}
and we have omitted the one term that appears to order $1/W^2$.

How do we decide which states should go into the model space? In order to keep the perturbation consistently of order $1/W$ we should, in principle, keep all states with nonzero amplitude onto $V\ket{n}$ in the model space where $\ket{n}$ is the state chosen initially. However, the expression for the matrix element simplifies matters. We consider operators ${O}$ that are either diagonal in the configuration basis or which flip a pair of spins amounting to a single term in $V$. Then we are justified in restricting the model space to $\ket{n}$ and ${O}\ket{n}$ if the latter lies within an energy window of $1/W$ from $\ket{n}$.

\subsection{Second Order Perturbation Theory}

We now discuss the perturbative corrections to eigenstates to second order in $V$. The first step in constructing the eigenstates is to include terms to second order in the effective Hamiltonian. 

\begin{align}
	\langle m \vert H^{(2)}_{\rm eff} \vert m'\rangle = \sum_{k,k'\in \overline{\text{D}}} \langle m \vert H \vert k \rangle \frac{ \langle k \vert V \vert k' \rangle \langle k' \vert V \vert m' \rangle}{ (E_0^{m'} - E_0^k) (E_0^{m'}-E_{0}^{k'}) } \nonumber \\ - \sum_{\tiny\begin{array}{c} k\in \overline{\mathrm{D}} \\  l \in \mathrm{D} \end{array}} \langle m \vert H \vert k \rangle \frac{ \langle k \vert V \vert l \rangle \langle l \vert V \vert m' \rangle}{ (E_0^{l} - E_0^k) (E_0^{m'}-E_{0}^{k}) }.
	\label{eq:Heff2}
	\end{align}

As above, the total effective Hamiltonian up to second order is given by the sum of all lower order contributions as 
\begin{equation}
H_\text{eff} = H_\text{eff}^{(0)} + H_\text{eff}^{(1)} + H_\text{eff}^{(2)}.
\end{equation}
Once we have computed the eigenstates of the effective Hamiltonian to this order, we again lift them into the full Hilbert space by operating on them with $\Omega^{(0)}+\Omega^{(1)}+\Omega^{(2)}$. The first two terms are reported above, the second order contribution is:

\begin{equation}
\begin{split}
\Omega^{(2)}\ket{v_\lambda} &= \sum_{\tiny\begin{array}{c} k,k'\in \overline{\mathrm{D}} \\  m \in \mathrm{D} \end{array}}
\ket{k} \frac{ \langle k \vert V \vert k' \rangle \langle k' \vert V \vert m \rangle}{ (E_0^{m} - E_0^k) (E_0^{m}-E_{0}^{k'}) } v^{\lambda}_m  \nonumber \\ &- \sum_{\tiny\begin{array}{c} k\in \overline{\mathrm{D}} \\  l,m \in \mathrm{D} \end{array}} \ket{k} \frac{ \langle k \vert V \vert l \rangle \langle l \vert V \vert m \rangle}{ (E_0^{l} - E_0^k) (E_0^{m}-E_{0}^{k}) }
v^{\lambda}_m.
\end{split}
\end{equation}

Diagonal matrix elements may now be computed. In full,

\begin{align}
& \langle \Psi_\lambda \vert {O} \ket{\Psi_\lambda} = \sum_{m,m' \in\mathrm{D}} v_m^{\lambda} v_{m'}^{\lambda \hspace{0.5mm}*} \langle m' \vert {O} \vert m \rangle \nonumber \\
& +  \sum_{\tiny\begin{array}{c} k\in \overline{\mathrm{D}} \\  m,m' \in \mathrm{D} \end{array}} v_m^{\lambda} v_{m'}^{\lambda \hspace{0.5mm}*} \langle m' \vert {O} \vert k \rangle \frac{ \langle k \vert V\vert m\rangle}{E_0^m - E_0^k} \nonumber \\
& +  \sum_{\tiny\begin{array}{c} k\in \overline{\mathrm{D}} \\  m,m' \in \mathrm{D} \end{array}} v_m^{\lambda} v_{m'}^{\lambda \hspace{0.5mm}*} \langle k \vert {O} \vert m \rangle \frac{ \langle m' \vert V\vert k\rangle}{E_0^{m'} - E_0^k} \nonumber \\
& +  \sum_{\tiny\begin{array}{c} k,k' \in \overline{\mathrm{D}} \\  m,m' \in \mathrm{D} \end{array}}  v_m^{\lambda} v_{m'}^{\lambda \hspace{0.5mm}*} \langle k' \vert {O} \vert k \rangle \frac{  \langle k \vert V \vert m \rangle \langle m' \vert V \vert k' \rangle }{(E_0^{m} - E_0^k) (E_0^{m'}-E_{0}^{k'})} \nonumber \\
& +  \sum_{\tiny\begin{array}{c} k,k' \in \overline{\mathrm{D}} \\  m,m' \in \mathrm{D} \end{array}}  v_m^{\lambda} v_{m'}^{\lambda \hspace{0.5mm}*} \langle m' \vert {O} \vert k \rangle \frac{  \langle k \vert V \vert k' \rangle \langle k' \vert V \vert m \rangle }{(E_0^{m} - E_0^k) (E_0^{m}-E_{0}^{k'})} \nonumber \\
& -  \sum_{\tiny\begin{array}{c} k \in \overline{\mathrm{D}} \\  l,m,m' \in \mathrm{D} \end{array}}  v_m^{\lambda} v_{m'}^{\lambda \hspace{0.5mm}*} \langle m' \vert {O} \vert k \rangle \frac{  \langle k \vert V \vert l \rangle \langle l \vert V \vert m \rangle }{(E_0^{l} - E_0^k) (E_0^{m}-E_{0}^{k})} \nonumber \\
& +  \sum_{\tiny\begin{array}{c} k,k' \in \overline{\mathrm{D}} \\  m,m' \in \mathrm{D} \end{array}}  v_{m'}^{\lambda} v_{m}^{\lambda \hspace{0.5mm}*} \langle k \vert {O} \vert m' \rangle \frac{  \langle k' \vert V \vert k \rangle \langle m \vert V \vert k' \rangle }{(E_0^{m} - E_0^k) (E_0^{m}-E_{0}^{k'})} \nonumber \\
& -  \sum_{\tiny\begin{array}{c} k \in \overline{\mathrm{D}} \\  m,m',l \in \mathrm{D} \end{array}}  v_{m'}^{\lambda} v_{m}^{\lambda \hspace{0.5mm}*} \langle k \vert {O} \vert m' \rangle \frac{  \langle l \vert V \vert k \rangle \langle m \vert V \vert l \rangle }{(E_0^{l} - E_0^k) (E_0^{m}-E_{0}^{k})} \nonumber \\
\end{align}

\section{Energy dependence of local operators}
\label{sec:e_dependence_loc_op}

\begin{figure}[htpb!]
	\centering
	\includegraphics[width=\columnwidth]{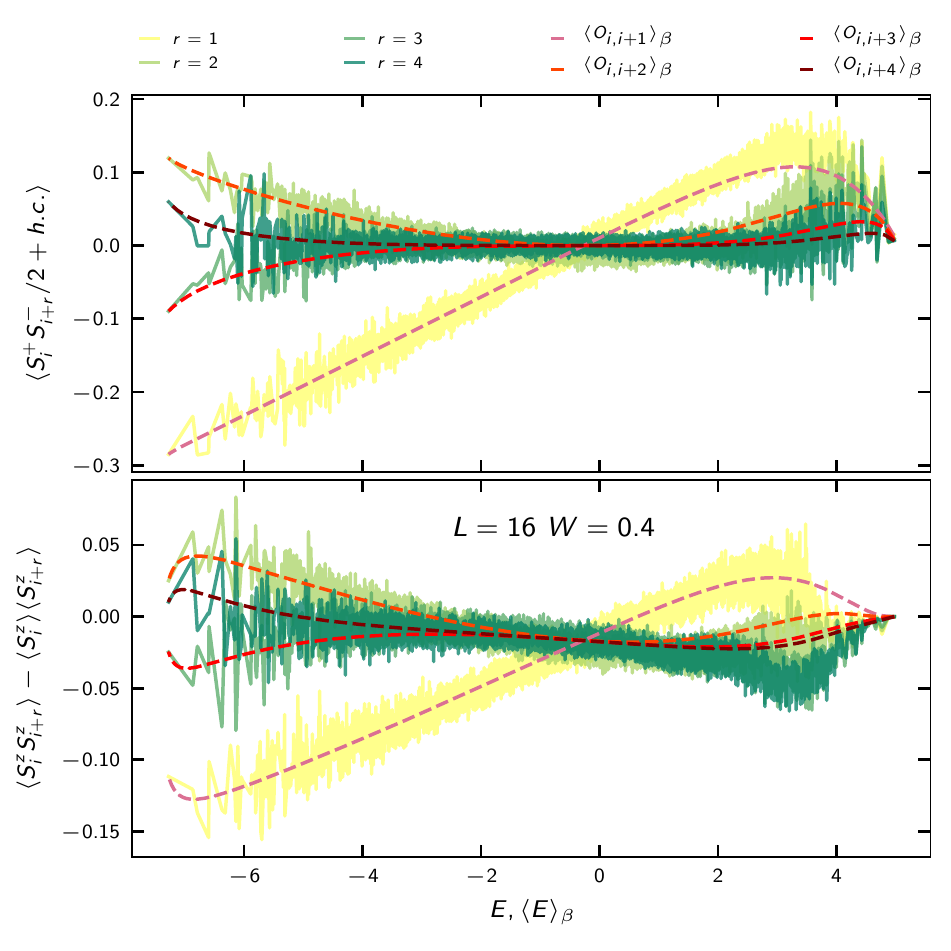}
	\caption{Spin-flip correlator and connected $z$ correlation as function of energy at
          distance $r=1,2,3,4$ in the ETH phase. Both correlations increase linearly with energy at
          distance $r=1$. The spin-flip correlator is constant at larger distances $r>1$ whereas the
          $z$ correlation goes down slightly. The dashed lines are the thermal expectation values in the canonical $M=0$ ensemble of the correlators $\langle O_{i,i+r} \rangle_\beta = \tr \left( \frac{\E^{-\beta H}}{Z} O_{i,i+r} \right)$ plotted vs. the expectation value of the energy $\langle E \rangle_\beta = \tr \left( \frac{\E^{-\beta H}}{Z} H \right)$ for $\beta\in[-1000,1000]$. }
	\label{fig:operators_vs_energy}
\end{figure}

In the ergodic regime, the validity of the ETH implies that energy eigenstate expectation values of local operators are equivalent to the thermal average at the temperature corresponding to the energy eigenvalue:
\begin{eqnarray}
	\bra{n} {O} \ket{n} \approx \tr \left( \dfrac{e^{-\beta (E_n) {H}}}{Z} {O}\right),
\end{eqnarray}
where $E_n$ is the energy density of the state $\ket{n}$ and $\beta(E_n)$ is the temperature chosen such that 
\begin{equation}
\tr \left( \frac{\E^{-\beta(E_n) H}}{Z}  H\right) = E_n.
\end{equation}
 In Fig. \ref{fig:operators_vs_energy} the expectation value of the spin-flip correlator and the connected $z$ correlation at distance $r=1,2,3,4$ are plotted as function of energy. The corresponding thermal average matches the mean value of $\bra{n} {O} \ket{n}$ .

We note that the thermal expectation values correspond to the mean of the distributions we consider in the main text. Since the correlators for $r=1$ are terms of the Hamiltonian, they exhibit a significant slope in the middle of the spectrum, leading to the observed sensitivity to the energy target. In the $M=0$ sector $\tr(H_{M=0})/\text{dim}(\mathcal{H}_{M=0})=-L/(4L-4) $ which sets the infinite temperature limit.

\section{Heavy tails in the MBL phase}
\label{sec:tails}
\begin{figure}[htbp!]
	\centering
	\includegraphics[width=\columnwidth]{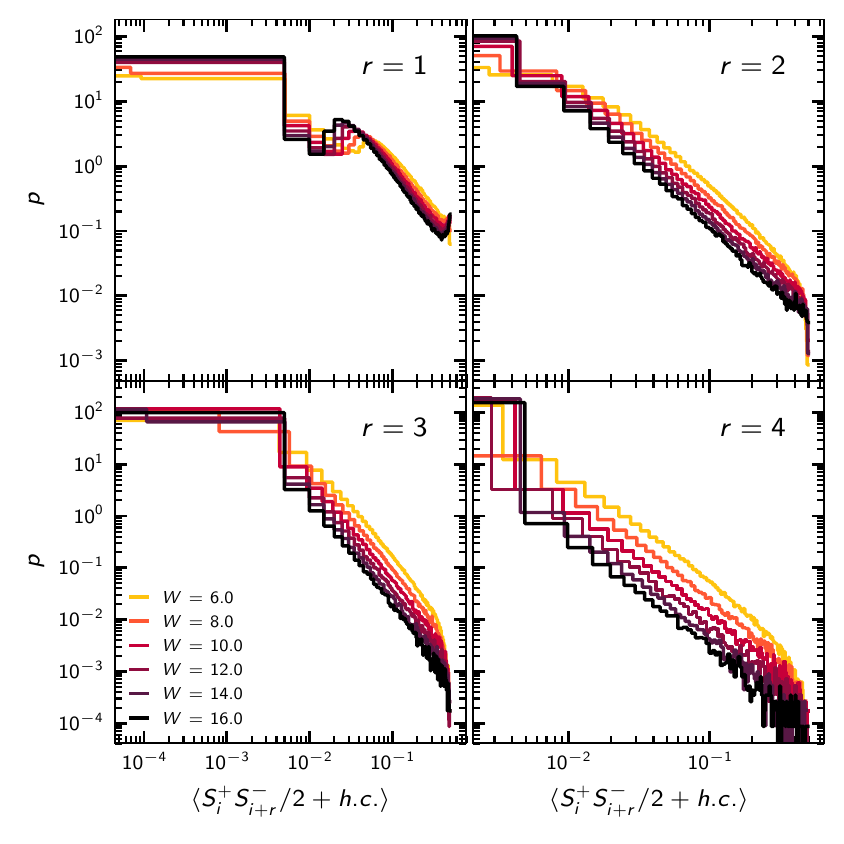}
	\caption{Probability distribution of the spin-flip correlator $\bra{n} {S}^+_i {S}^-_{i+r}/2+h.c. \ket{n} $ at distance $r=1,2,3,4$ for system size $L=20$ in the MBL phase. Putting aside the special features at short distance $r=1$, sharp peaks at zero and power law tails elsewhere are the common feature. The disorder strength dependence seems to be stronger at large distances.  }
	\label{SxSy_tails}
\end{figure}

\begin{figure}[htbp!]
	\centering
	\includegraphics[width=\columnwidth]{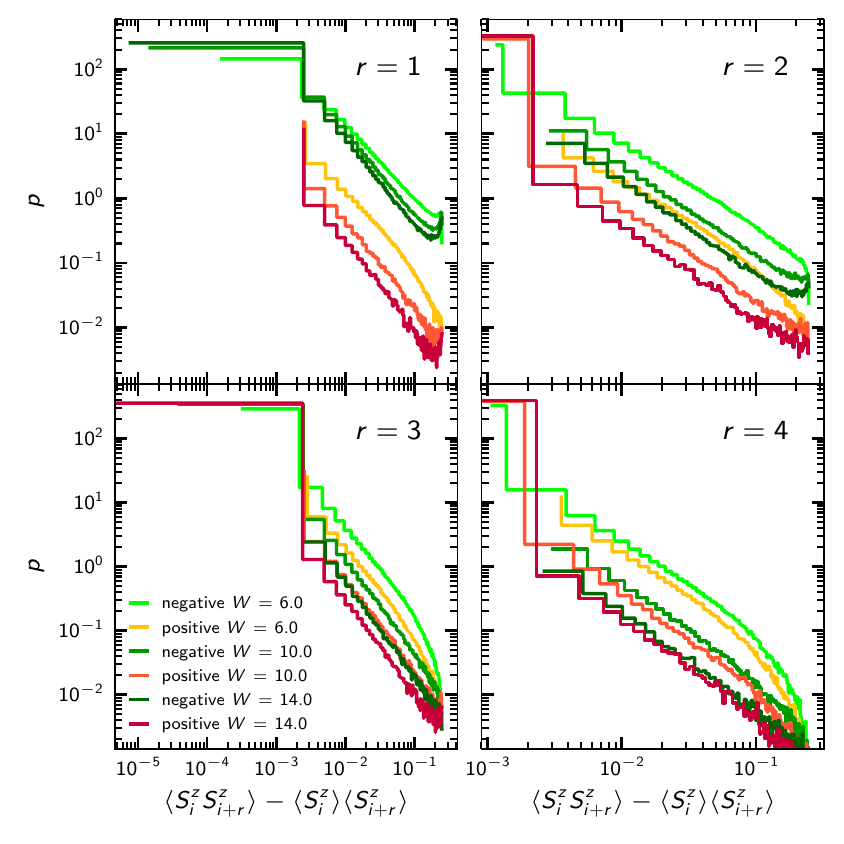}
	\caption{Probability distribution of correlation function $\bra{n} S^z_i S^z_{i+r} \ket{n} - \bra{n} S^z_i \ket{n} \bra{n} S^z_{i+r} \ket{n}$ at distances $r=1,2,3,4$ for system size $L=20$ in the MBL phase. The green and yellow-red curves correspond to negative and positive values of the correlation respectively. As seen in the main text, these distributions are asymmetric around zero, though they seems to have the same power law behavior on both sides. }
	\label{SzSz_tails}
\end{figure}

Some common features of the matrix elements distributions of correlation functions in the MBL phase are the sharp peaks at zero and the presence of tails. In Fig. \ref{SxSy_tails} and \ref{SzSz_tails} such tails are highlighted for different distances using a doubly logarithmic scale. 
They both show heavy tails, which seem consistent with a power law behavior. 
Since the connected $S_i^zS_{i+r}^z$ correlator is asymmetric, we show the positive part of the distribution separately from the negative part in Fig. \ref{SzSz_tails} (the sign of the negative part is flipped to show them on the same plot).

Remarkably, the right and left hand tails of the connected $z$ correlation seem to follow the same power law dependence despite their asymmetry, which is clearly visible in this representation.

\end{document}